\documentclass[twocolumn,nofootinbib,showpacs,preprintnumbers]{revtex4}
\usepackage{graphicx}
\usepackage{dcolumn}
\usepackage{bm}
\newcommand{\mcur}{m_{\mbox{\scriptsize cur}}}
\newcommand{\mhcur}{\hat{m}_{\mbox{\scriptsize cur}}}
\newcommand{\mucur}{m_{u,\mbox{\scriptsize cur}}}
\newcommand{\mdcur}{m_{d,\mbox{\scriptsize cur}}}
\newcommand{\mscur}{m_{s,\mbox{\scriptsize cur}}}
\newcommand{\scur}{\mbox{\scriptsize cur}}
\newcommand{\mdyn}{m_{\mbox{\scriptsize dyn}}}
\newcommand{\mcon}{m_{\mbox{\scriptsize con}}}
\newcommand{\mhcon}{\hat{m}_{\mbox{\scriptsize con}}}
\newcommand{\mscon}{m_{s,\mbox{\scriptsize con}}}
\newcommand{\mcl}{m_{\mbox{\scriptsize CL}}}
\newcommand{\lsm}{L$\sigma$M}
\newcommand{\qllsm}{QLL$\sigma$M}
\newcommand{\etc}{\mbox{\scriptsize ETC}}
\newcommand{\nq}{\mbox{\scriptsize NQ}}
\newcommand{\sms}{\mbox{\scriptsize SS}}
\newcommand{\cl}{\mbox{\scriptsize CL}}
\newcommand{\chpt}{\mbox{\scriptsize ChPT}}
\newcommand{\hochpt}{\mbox{\scriptsize HOChPT}}
\newcommand{\gmo}{\mbox{\scriptsize GMO}}
\newcommand{\gmor}{\mbox{\scriptsize GMOR}}
\newcommand{\vmd}{\mbox{\scriptsize VMD}}
\newcommand{\chsb}{\mbox{\scriptsize ChSB}}
\newcommand {\bom}[1]{\boldmath{$#1$}}

\begin{document}

\title{Pion Chiral Symmetry Breaking in the\\Quark-Level Linear Sigma Model
and Chiral Perturbation Theory}

\author{Michael D.\ Scadron}
\email{scadron@physics.arizona.edu}
\affiliation{Physics Department, University of Arizona, Tucson,
AZ 85721, USA}

\author{Frieder Kleefeld}
\email{kleefeld@cfif.ist.utl.pt}
\author{George Rupp}
\altaffiliation[Corresponding author]{}
\email{george@ist.utl.pt}
\affiliation{Centro de F\'{\i}sica das Interac\c{c}\~{o}es Fundamentais (CFIF),
Instituto Superior T\'{e}cnico,
Av. Rovisco Pais,
P-1049-001 LISBOA,
Portugal}

\date{\today}

\begin{abstract}
Chiral symmetry breaking (ChSB) is reviewed to some extent within the
quark-level-linear-sigma-model (QLL$\sigma$M) theory and standard chiral
perturbation theory (ChPT). It is shown, on the basis of several examples
related to the pion, as a well-known Goldstone boson of chiral symmetry
breaking, that even the non-unitarized QLL$\sigma$M approach accounts, to
a good approximation, for a rather simple, self-consistent, linear, and
very predictive description of Nature.  On the other hand, ChPT --- even
when unitarized --- provides a highly distorted, nonlinear, hardly
predictive picture of Nature, which fits experiment only at the
price of a lot of parameters, and requires a great deal of unnecessary
theoretical effort. As the origin of this distortion, we identify the fact
that ChPT, reflecting only direct ChSB by nonvanishing, current-quark-mass
values, does not --- contrary to Quantum Chromodynamics (QCD) and the
QLL$\sigma$M --- contain any mechanism for the spontaneous generation of the
dynamical component of the constituent quark mass. This leads to a very
peculiar picture of Nature, since the strange current quark mass has to
compensate for the absence of nonstrange dynamical quark masses. We thus
conclude that standard ChPT  --- contrary to common wisdom --- is unlikely to
be the low-energy limit of QCD. On the contrary, a chiral perturbation
theory derived from the QLL$\sigma$M, presumably being the true low-energy
limit of QCD, is expected instead to provide a distortion-free description of
Nature, which is based on the heavy standard-model Higgs boson as well as
light scalar mesons, as the source of spontaneous generation of current and
dynamical quark masses, respectively.

\end{abstract}

\maketitle

\section{Introduction}
\label{sec1}
The fundamental laws of physics all respect some exact symmetries. In Nature,
however, we are mostly --- and fortunately -- facing symmetries which are
broken. In fact, if the electromagnetic,
weak, strong, and gravitational interactions obeyed exactly the same
symmetries, the observable world would be a dull place. It is precisely the
mismatch between the symmetries respected individually by the different
interactions that gives rise to the amazing diversity and beauty of Nature at
high, intermediate, and particularly low energies \cite{Eidelman:2004wy}.

In this paper we shall focus on the chiral symmetry underlying the theory of
strong interactions, as they interfere with chiral-symmetry-breaking (ChSB)
electroweak interactions. For convenience, our attention will be concentrated
on observables associated with the gold-plated test particle of ChSB, i.e.,
the pion \cite{Marques:2005}, which, being a Goldstone boson, would be
massless in the so-called chiral limit (CL), that is, the limit of strong
interactions without electroweak interactions.

As will be discussed in more detail below, a convenient quantity to discuss
and measure the amount of ChSB in the pion is not only its mass $m_\pi$, but
also the underlying (nonstrange) \textit{constituent quark mass} $\mcon$.
The latter can be additively decomposed as $\mcon=\mdyn+\mcur$
\cite{Elias:1984aw} into a bulk part called \textit{dynamical quark mass}
$\mdyn$ associated with chiral-symmetric strong interactions, and a small
correction called \textit{current quark mass} $\mcur$, being nonzero only
in the presence of ChSB electroweak interactions. Hence, in the CL we have
$m_\pi\rightarrow 0$, $\mcur\rightarrow 0$, $\mdyn\rightarrow\mcl$,
and $\mcon\rightarrow\mcl$, where $\mcl$ may be called
\textit{chiral-limiting quark mass}.

Thus, it becomes clear that a self-consistent picture of ChSB will only be
achieved through a complete understanding of electroweak and strong
interactions, especially their interplay.
Unfortunately, and maybe somewhat surprisingly, neither the  experimental
nor the theoretical situation is even close to settled, despite the seemingly
overwhelming success of the present-day standard model of particle physics
(SMPP), as we explain next.

In the first place, despite the consensus on the mechanism for generating the
tiny up/down current quark masses ($\sim10^1$ MeV) in the Lagrangian density
of the SMPP, the corresponding heavy scalar ($m_H>10^5$ MeV), the Higgs boson,
remains undetected.

Secondly, there is substantial experimental evidence \cite{Eidelman:2004wy} for
the existence of a complete nonet of light ($m<1$ GeV) scalar mesons, which can
be considered responsible for the dynamical generation of the surprisingly
large $\mdyn$ ($\sim300$ MeV). However, our present mastering of QCD
\cite{Muta:1987mz}, the commonly accepted theory underlying strong
interactions, does not allow to predict the existence of light scalar mesons,
not to speak of their multiplet structure and pole positions. But at
least one can gain, e.g.\ on the basis of QCD-inspired variational chiral quark
models \cite{Bicudo:1989sh}, Bethe-Salpeter equations (BSE)
\cite{Delbourgo:1982tv,Scadron:1982eg,LeYaouanc:1984dr,Miransky:1993}, or Dyson-Schwinger
equations (DSE)
\cite{Miransky:1993,Fischer:2004ym}, some understanding about the dynamical generation of
$\mdyn$, once small ChSB values for $\mcur$ are given. Unfortunately, most DSE
approaches are Euclidean, whereas it is not clear whether a Wick rotation from
Minkowski \cite{Sauli:2004bx,Bicudo:2003fd} to Euclidean space will yield
identical results.

Finally, chiral perturbation theory (ChPT) (see e.g.\ Ref.\
\cite{Donoghue:1992dd}), the mainstream low-energy \em effective \em approach
to strong interactions, is only capable of recovering the light scalar mesons
when it is unitarized \cite{Oller:1998hw,Caprini:2005zr}
by hand. However, its exact relation
to QCD remains unexplained, despite well-known claims made by the large ChPT
community. Moreover, ChPT relies upon a strongly distorted picture of Nature,
namely based on current quarks, whose masses are at least two orders of
magnitude smaller than those of most observed hadrons. As ChPT does not provide
any mechanism to generate $\mdyn$, contrary to QCD as well as chiral quark
models and BSE/DSE approaches, it is the strange current quark mass, taken at
roughly 25 times the nonstrange current mass, which partly has to take
over the role of $\mdyn$. This, we believe, is the origin, of, for example,
the excessively large strangeness content of the proton predicted by ChPT, and,
as discussed below, the unacceptably large value for
$\sigma^{\bar{s}s}_{\pi N}$ ($\simeq 10$~MeV), which has been measured to be
$\le 2$~MeV \cite{Foudas:1989rk}. Furthermore, an additional and very sizable
amount of glue, which logically should already be accounted for in ChPT as
the supposed ``low-energy limit of QCD'', is still needed e.g.\ to explain the
nucleon mass \cite{glue}.

The aim of the present paper is to show that there exists an alternative
approach to strong interactions, much simpler than ChPT and free from its
distortions and shortcomings, yet highly predictive and also intuitively
appealing, namely the quark-level linear $\sigma$ model (\qllsm). The
original \lsm, with nucleons instead of quarks as the fermionic fields, had
been suggested by Schwinger \cite{Schwinger:1957em}, and was then proposed
by Gell-Mann and L\'{e}vy \cite{Gell-Mann:1960np} as the preferable model for
strong interactions featuring spontaneous ChSB, in particular respecting
partial conservation of axial-vector currents (PCAC). On the other hand, in the
same epoch Nambu and Jona-Lasinio (NJL) \cite{Nambu:1961tp} presented a
nonrenormalizable model for the pion, originally encompassing (anti)nucleons
but straightforwardly extensible to quarks, which was based on dynamical ChSB.
Then, the \lsm\ was generalized to quarks by L\'{e}vy \cite{Levy:1967a} and
Cabibbo \& Maiani \cite{Cabibbo:1970uc}. Finally, Delbourgo and Scadron
\cite{Delbourgo:1993dk} managed to dynamically generate the \qllsm\
self-consistently in one-loop order, which considerably increased its
predictive power.  For instance, the famous chiral-limiting NJL result
$m_\sigma=2\mcl$ is reproduced, i.e., in the chiral limit the mass of the
lightest scalar meson equals twice the dynamical quark mass, with the pion
being massles, of course, in the same limit.

Clearly, the simplicity and linearity of the \qllsm\ is owing to the
inclusion of the $\sigma$ meson, now experimentally confirmed
\cite{Eidelman:2004wy}, as an elementary field being the chiral partner of the
pion \cite{vanBeveren:2002mc,Scadron:2003yg}
whereby in loop order both are self-consistently recovered as $0^{-+}$ 
and $0^{++}$ $\bar{q}q$ states, respectively. In contrast, ChPT requires
unitarization, i.e., infinite order, to resuscitate the $\sigma$
\cite{Oller:1998hw,Caprini:2005zr}. As we shall demonstrate through several examples, the
\qllsm\ is doing a much better job, even without unitarization, in reproducing
a large variety of low-energy observables than ChPT, with only a tiny fraction
of the effort needed. Finally, we shall present arguments why the \qllsm\ may
even constitute a full-fledged, asymptotically free theory of strong
interactions.

In the present paper, we shall pass in review the mechanism of ChSB for a
variegation of low-energy observables and relations, employing the \qllsm\ as
compared to standard ChPT, namely:
pion and nucleon sigma terms (Sec.~II), meson charge radii (Sec.~III), 
Goldberger-Treiman relations (Sec.~IV), Goldberger-Treiman discrepancies
(Sec.~V), constituent quark mass via baryon magnetic moments (Sec.~VI),
effective current quark and ChSB pion, kaon, $\eta_8$ masses (Sec.~VII),
ground-state scalar $\bar{q}q$ nonet (Sec.~VIII), $I=0$ $\pi\pi$ scattering
length (Sec.~IX), and the process $\gamma\gamma\to\pi^0\pi^0$ (Sec.~X).
A summary with discussion and conclusions is presented in Sec.~XI.

\section{\bom{\sigma_{\pi\pi}} and \bom{\sigma_{\pi N}} chiral breaking terms}
\label{sec2}
The pion $\sigma$ term is considered a $c$-number ($t$-independent). It is
defined at $q^2=t=m^2_\pi$ via PCAC as (with $f_\pi\simeq 93$ MeV for $\pi^0$,
the Particle Data Group (PDG) 2004 \cite{Eidelman:2004wy} takes $f_\pi\simeq
(92.42 \pm 0.26)$ MeV; ETC stands for ``Equal-Time Commutator'')
\begin{eqnarray} \sigma_{\pi\pi} \, \delta_{ij} & = &
\left<\pi_i\right|[Q^5_{\pi}, i \partial A^\pi ]_{\etc}\left|\pi_j\right>
\nonumber \\
& \simeq & \frac{1}{f_\pi} \left<0\right|\partial A^\pi_{i} (0)
\left|\pi_j\right> = m^2_\pi  \, \delta_{ij}\; . \label{eqsigpipi1}
\end{eqnarray}
The analogue $\pi N$ $\sigma$ term is defined as \cite{Scadron:1997cs}
\begin{equation} \sigma_{\pi N} \, \bar{u}_N u_N \; = \;
\left<N\right|[Q^5_{\pi 3}, i \partial A_3^\pi ]_{\etc} \left|N\right>
\; ,\label{eqsigpin1}
\end{equation}
also taken as a $c$-number in the 1960's.

Both the nonvanishing values of Eqs.~(\ref{eqsigpipi1}) and (\ref{eqsigpin1})
characterize $SU(2)\times SU(2)$ ChSB.

\begin{figure*}
\includegraphics{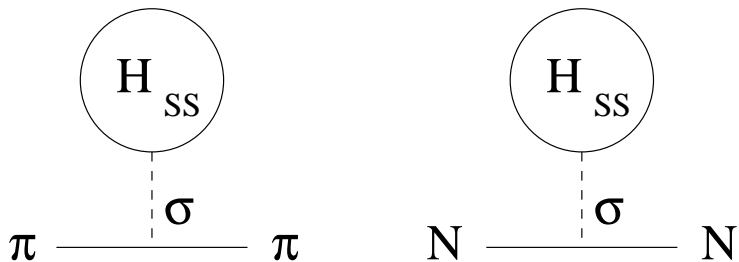}
\caption{\label{fig:tadpoles1} Tadpoles generated by the ChSB ``semi-strong''
(SS) Hamiltonian $H_{\sms}$.}
\end{figure*}

In the language of the \lsm, Eqs.~(\ref{eqsigpipi1}) and (\ref{eqsigpin1}) are
represented by ``nonquenched'' (NQ) $\sigma$ tadpole graphs, shown in
Fig.~\ref{fig:tadpoles1}, with \lsm\ couplings $2 g_{\sigma \pi\pi} =
(m^2_\sigma - m^2_\pi)/f_\pi\simeq m^2_\sigma/f_\pi$, $g_{\sigma NN}
\simeq m_N/f_\pi$. 
Both tadpole ``heads'' are generated by the ChSB Hamiltonian ($H_{\sms}$ means
semi-strong Hamiltonian) implying
\begin{equation} \sigma^{\nq}_{\pi N} = \left( \frac{m_\pi}{m_\sigma}\right)^2
m_N \simeq 40 \; \mbox{MeV} \; ,
\end{equation}
for an averaged nucleon mass $m_N= 938.9$~MeV and chiral-limiting $\sigma$ mass
$m^{\cl}_\sigma$ of $650.8$~MeV (see Sec.~\ref{sec8} and
Ref.~\cite{Delbourgo:1993dk}), predicting an ``on-shell'' broad $\sigma$ mass
of \cite{Delbourgo:1998kg,Surovtsev:2002kr,Nagy:2004tp}
$m_\sigma\simeq 665$~MeV, for $(m^{\cl}_\sigma)^2 \simeq m^2_\sigma - m^2_\pi$.

The perturbative ``quenched'' $\pi N$ $\sigma$ term found by GMOR
\cite{Gell-Mann:1968rz}, or via the APE collaboration \cite{Cabasino:1991mr},
is about
\begin{eqnarray}
\sigma^{\gmor}_{\pi N} & = & \frac{m_{\Xi} + m_\Sigma -
2 m_N}{2} \left( \frac{m^2_\pi}{m^2_K-m^2_\pi} \right) \nonumber \\
 &\simeq & 26 \; \mbox{MeV} \; ,
\end{eqnarray}
so the net $\pi N$ $\sigma$ term in the L$\sigma$M is the sum
\begin{equation}
\sigma^{L\sigma M}_{\pi N} = \sigma^{\gmor}_{\pi N} + \sigma^{\nq}_{\pi N}
\simeq (26 + 40) \; \mbox{MeV} =  66 \; \mbox{MeV}\; .
\label{eqgmor1}
\end{equation}

While the Adler--Weinberg \cite{Adler:1964um,Weinberg:1966kf}
$\pi N$ low-energy theorems have
large background terms, the Cheng--Dashen (CD) \cite{Cheng:1970mx} theorem at
$t=2m^2_\pi$ has a very small background term (for the isospin-even $\pi N$
amplitude), viz.\
\begin{equation} \overline{F}^+(\nu=0,t=2m^2_\pi) =
\frac{\sigma_{\pi N}}{f^2_\pi} + {\cal O}(m^4_\pi) \simeq 1.020\;m^{-1}_\pi \; ,
\label{eqcd1}
\end{equation}
as measured by Koch and H\"ohler \cite{Koch:1982pu}, corresponding to
(for $f_\pi \simeq 93$~MeV)
\begin{equation}
\sigma_{\pi N} = (1.018) \frac{f^2_\pi}{m_{\pi^+}} \simeq 63 \; \mbox{MeV} \; ,
\label{eqhoehler1}
\end{equation}
where the small ${\cal O}(m^4_\pi)$ CD term in Eq.\ (\ref{eqcd1}) is
$0.002\,m^{-1}_\pi$.

Or instead one can work in the infinite-momentum frame (IMF), which suppresses
the tadpoles and predicts \cite{Clement:1990kj} (requiring squared baryon
masses with cross term $2 m_N \sigma_{\pi N}$)
\begin{equation}
\sigma_{\pi N} = \frac{m^2_{\Xi} + m^2_\Sigma - 2 m^2_N}{2 m_N}
\left( \frac{m^2_\pi}{m^2_K-m^2_\pi} \right)\simeq 63 \; \mbox{MeV} \; .
\label{eqimf1}
\end{equation}
Note that the data in Eq.~(\ref{eqhoehler1}) or the IMF value in
Eq.~(\ref{eqimf1}) of $63$~MeV are compatible with the L$\sigma$M value in
Eq.~(\ref{eqgmor1}) of $66$~MeV. 

However, ChPT leads to a more complicated scheme. By 1982, the review by Gasser
and Leutwyler preferred $\sigma_{\pi N} \simeq 24$ to $35$~MeV
\cite{Gasser:1982ap}. In 1984, their revised ChPT \cite{Gasser:1983yg} ruled out
some L$\sigma$M in Appendix~B \em ``$\ldots$ as a realistic alternative to QCD
$\ldots$'', \em and the ``scalar form factor'' ChPT scheme of 1990
\cite{Donoghue:1990xh} used \textit{doubly-subtracted} \/dispersion relations
(adding 6 new paramters) via a low-mass-Higgs decay $H\rightarrow \pi\pi$, yet
to be measured. Then, in 1991, ChPT obtained \cite{Gasser:1990ce} a
non-$c$-number $\pi N$ $\sigma$ term with 
\begin{equation} \sigma_{\pi N} (t=0) \simeq 45 \; \mbox{MeV} \; ,
\label{eqsigterm1991} 
\end{equation}
and a larger $\pi N$ $\sigma$ term at $t=2m^2_\pi$  of $60$~MeV as
(HOChPT stands for ``higher-order ChPT'')
\cite{Leutwyler:1992ry} (in MeV)
\begin{eqnarray} \sigma_{\pi N}(2m^2_\pi) & = & \sigma^{\gmor}_{\pi N} (25) +
\sigma^{\hochpt}_{\pi N} (10) \nonumber \\
& + & \sigma^{\bar{s}s}_{\pi N} (10) +
\sigma^{t\mbox{\scriptsize-dep.}}_{\pi N} (15) \nonumber \\
& \simeq & 60 \; \mbox{MeV} \; , \label{eqsigtermchpt1} 
\end{eqnarray}
because the latter ``three pieces {\em happen} to have the same sign as
$\sigma^{\mbox{\scriptsize GMOR}}_{\pi N}$'' \cite{Leutwyler:1992ry}. Note
that the $\sigma^{\bar{s}s}_{\pi N}$ term has been measured as $\le 2$~MeV
\cite{Foudas:1989rk}. Such a ``happening'' suggests that ChPT in
Eq.~(\ref{eqsigtermchpt1}) is physically less significant and much more
complicated than Eqs.~(\ref{eqgmor1}), (\ref{eqhoehler1}), or (\ref{eqimf1})
above.

\section{Meson charge radii}
 \label{sec3}
Using vector and axial-vector form factors, the PDG tables on pages 499 and 621
of Ref.~\cite{Eidelman:2004wy} state that the pion and kaon charge radii are
measured as
\begin{eqnarray} r_{\pi^+} & = & (0.672\pm0.008)\;\mbox{fm}\; ,
\label{eqrchargeexp1} \\
r_{K^+} & = & (0.560\pm0.031)\;\mbox{fm}\; .
\end{eqnarray}
In fact, the vector-meson-dominance (VMD) and L$\sigma$M schemes give nearby
values \cite{Scadron:2002mm} (with $\hat{m}\equiv (m_u+m_d)/2$ and $\hbar c
\simeq 197.3$~MeV$\,$fm)
\begin{eqnarray} r^{\vmd}_{\pi} & = & \makebox[2.4cm][l]{$\sqrt{6} / m_\rho$} \;
\simeq \; 0.623\;\mbox{fm}\; , \label{eqrpicharge1} \\
r^{\mbox{\scriptsize L$\sigma$M}}_{\pi} & = &
\makebox[2.4cm][l]{$1/\hat{m}_{\cl}$} \; \simeq \; 0.63\;\mbox{fm}\; ,
\label{eqrpicharge2} \\
r^{\vmd}_{K} & = & \makebox[2.4cm][l]{$\sqrt{6} / m_{K^\ast}$} \; \simeq \;
0.54\;\mbox{fm}\; , \\
r^{\mbox{\scriptsize L$\sigma$M}}_K & = &
\makebox[2.4cm][l]{$2/(\hat{m}+m_s)_{\cl}$} \; \simeq \; 0.51\;\mbox{fm}\; .
\end{eqnarray}
Note that $\hat{m}_{\cl}\simeq 313$~MeV as we shall point out later in
Sec.~\ref{sec6}. Then the $\bar{q}q$ pion is very tightly bound, almost a
fused $\bar{q}q$ meson.

Moreover, on page 498 of Ref.~\cite{Eidelman:2004wy}, the pion vector and
axial-vector form factors are observed at $q^2=0$ as
\begin{eqnarray} F^\pi_V (0) & = & 0.017\pm 0.008 \; , \label{eqvecff1} \\
 F^\pi_A (0) & = & 0.0116\pm 0.0016 \; .\label{eqaxff1}
\end{eqnarray} 
Using $f_\pi\simeq 93$~MeV, the theoretical CVC estimates of these form factors
are \cite{VI58} (see also Ref.~\cite{Scadron:2002mm})
\begin{eqnarray} F^\pi_V (0) & = & \frac{m_{\pi^+}}{8\pi^2f_\pi}\;\,\; \simeq
\; 0.0190 \; , \label{eqvecff2} \\
 F^\pi_A (0) & = & \frac{m_{\pi^+}}{12\pi^2f_\pi}\; \simeq \; 0.0127 \; ,
\label{eqaxff2}
\end{eqnarray} 
reasonably near the data in Eqs.\ (\ref{eqvecff1}) and (\ref{eqaxff1}).

Sakurai's \cite{Sakurai:1973} vector-meson-coupling universality (VMU)
\cite{Djukanovic:2004mm} suggests $g_\rho \approx g_{\rho\pi\pi} \sim 6$,
whereas the L$\sigma$M predicts \cite{Bramon:1997gg} that VMU receives a
correction of $\frac{1}{6}\, g_{\rho\pi\pi}$ via the mesonic $\pi\sigma\pi$
loop, giving $g_{\rho\pi\pi}=g_\rho+\frac{1}{6}\, g_{\rho\pi\pi}$ or
\begin{equation} \frac{g_{\rho\pi\pi}}{g_\rho} \; = \; \frac{6}{5} \; ,
\end{equation}
close to the data \cite{Eidelman:2004wy,Scadron:2002mm} 
\begin{equation} g_{\rho\pi\pi} \simeq 5.95\; , \; g_\rho \simeq 4.96 \; .
\label{eqcoup1} \end{equation}
These couplings in (\ref{eqcoup1}) follow from the newly measured decay rates
\cite{Eidelman:2004wy} $\Gamma(\rho\pi\pi)=150.3$~MeV and
$\Gamma(\rho ee)=7.02$~keV (neglecting errors), as in Eqs.\ (11) and (12) of
Ref.~\cite{Scadron:2002mm}.

Although VMD and the L$\sigma$M reasonably match charge-radii data, the
one-loop-order ChPT prediction gives \cite{Donoghue:1989sq}
\begin{equation} \left< r^2_\pi\right> \; = \; 12 L_9/f^2_\pi \, ,
\end{equation}
which does \textit{not} \/uniquely predict data, since $L_9$ is a ChPT
low-energy-constant (LEC) parameter.

Whereas ChPT does not explain the pion or kaon charge radii, the L$\sigma$M
continues to match data.

Furthermore, two-loop ChPT \cite{Colangelo:2001sp,Ananthanarayan:2004xy}
suggests that the pion scalar-form-factor radius obeys
\begin{equation} \left< r^2_S\right> \; = \; (0.61 \pm 0.04)\;\mbox{fm}^2 \, .
\label{eqscalrad1}
\end{equation}
This value is debated \cite{Yndurain:2005cm,Pelaez:2003eh} as
\begin{equation} \left< r^2_S\right> \; = \; (0.75 \pm 0.07)\;\mbox{fm}^2 \, .
\label{eqscalrad2}
\end{equation}
We in turn claim that the average of Eqs.~(\ref{eqscalrad1}) and
(\ref{eqscalrad2}) is 75\% greater than the square of the VMD and L$\sigma$M
pion charge radius values in Eqs.~(\ref{eqrpicharge1}) and
(\ref{eqrpicharge2}), the latter two being compatible with data in
Eq.~(\ref{eqrchargeexp1}). Moreover, the scalar-form-factor basis of
Eqs.~(\ref{eqscalrad1}) and (\ref{eqscalrad2}) has \textit{neither} \/been
verified experimentally, \textit{nor} \/theoretically via VMD or the
L$\sigma$M, in contrast with the vector \& axial-vector form factors in
Eqs.~(\ref{eqvecff1}--\ref{eqaxff2}), or
Eqs.~(\ref{eqvecaxffsum1},\ref{eqvecaxffsum2}) to follow.

If we followed the discussion of Gasser and Leutwyler in
Ref.~\cite{Gasser:1983yg}, the identity
\begin{equation} \frac{f_\pi(m^2_\pi)}{f_\pi(0)} - 1 \; =\; \frac{1}{6} \,
m^2_\pi \left< r^2_S\right>  + \frac{13}{192\,\pi^2} \frac{m^2_\pi}{f^2_\pi}
\, + \, {\cal O}(m^4_\pi) \label{eqscalrad3}
\end{equation}
would hold. From Eq.~(\ref{eqdisprel1}) in the upcoming Sec.~\ref{sec5}, we
know that, to a good approximation,
\begin{equation} \frac{f_\pi(m^2_\pi)}{f_\pi(0)} - 1 \; =\;
\frac{m^2_\pi}{8\pi^2 f^2_\pi}  \, + \, {\cal O}(m^4_\pi)\; . 
\end{equation}
This would imply, in combination with Eq.~(\ref{eqscalrad3}) and
$\left< r^2_\pi\right>\simeq 1/\hat{m}^2 = 1/(g f_\pi)^2
\simeq N_c/(4 \, \pi^2 f^2_\pi)$, that
\begin{eqnarray} \left< r^2_S\right> & = & \frac{6}{m^2_\pi} \left(
\frac{m^2_\pi}{8\pi^2 f^2_\pi}  \, - \,\frac{13}{192\,\pi^2}
\frac{m^2_\pi}{f^2_\pi} \right) \nonumber \\
 & = & \left( 1-\frac{13}{24} \right) \, \frac{3}{4 \, \pi^2 f^2_\pi} \;
\simeq \; \frac{11}{24} \, \left< r^2_\pi\right> \; \simeq \;
0.19\;\mbox{fm}^2 \; , \nonumber \\
\end{eqnarray}
which is clearly in strong conflict with the ChPT results
Eqs.~(\ref{eqscalrad1}) and (\ref{eqscalrad2}).

In order to extend the L$\sigma$M to $SU(3)$, the $K^+\rightarrow e^+\nu\gamma$
form factor \textit{sum} \/is measured, according to page 621 in
Ref.~\cite{Eidelman:2004wy}, as
\begin{equation} \left| F^K_V (0)+F^K_A (0)\right| \; = \; 0.148\pm 0.010 \; .
\label{eqvecaxffsum1}
\end{equation}
The $SU(3)$ L$\sigma$M pole terms add up to \cite{Bramon:1992vp,Scadron:2002mm}
\begin{equation} \left| F^K_V (0)+F^K_A (0)\right|_{\mbox{\small L$\sigma$M}}
\; = \; 0.109 + 0.044 \; = \; 0.153 \; ,
\label{eqvecaxffsum2}
\end{equation}
compatible with data in Eq.~(\ref{eqvecaxffsum1}). The quark-loop prediction
for the pion charge radius was initially found in Refs.~\cite{Tarrach:1979ta}
via quark (and meson) loops in a L$\sigma$M framework, but \textit{not} \/via
ChPT. The above-measured vector and axial-vector form factors in
Eqs.~(\ref{eqvecff1}), (\ref{eqaxff1}), and (\ref{eqvecaxffsum1}) have
\textit{not} \/been extended to the mentioned ChPT scalar form factors, as was
suggested in deriving Eq.~(\ref{eqsigterm1991}), and will be suggested in
Eqs.~(\ref{eqgtdiscrep1}) and the nonlinear $a_{00}$ in Secs.~\ref{sec5},
\ref{sec9}, respectively.
\section{Goldberger-Treiman relations}
\label{sec4}
First we note that the pion-quark coupling in the QLL$\sigma$M is
\cite{Delbourgo:1993dk} $g_{\pi qq}=2\pi/\sqrt{3} \simeq 3.6276$, also
following from infrared QCD \cite{Elias:1984zh} and the $Z=0$ compositeness
condition \cite{Salam:1962ap}. Then the pion Goldberger-Treiman relation (GTR)
gives the nonstrange constituent quark mass as
\begin{equation}
f_\pi g \; = \; \hat{m} \; = \; 93\; \mbox{MeV} \times
\frac{2\pi}{\sqrt{3}} \; \simeq \; 337.4 \; \mbox{MeV} \, .
\label{eqgtr1}
\end{equation}
The nonrelativistic quark model predicts \cite{Eidelman:2004wy}, via magnetic
dipole moments (see Sec.~\ref{sec6}, Eq.~(\ref{mcon})), practically the same
value, viz.\
\begin{equation}
\hat{m} \; = \; \frac{1}{2} (m_u+m_d) \; \simeq \; 337.5 \; \mbox{MeV} \; ,
\label{eqmhat1}
\end{equation}
with $m_d-m_u\simeq 4$~MeV following from the kaon
\cite{Eidelman:2004wy,Delbourgo:1998qe} mass difference or the
$\Sigma^--\Sigma^+$ baryon mass difference, yielding the constituent masses
\begin{equation}
m_u \; \simeq \; 335.5 \; \mbox{MeV} \;\;\; , \;\;\; 
m_d \; \simeq \; 339.5 \; \mbox{MeV} \; .
\end{equation}

As for the strange quark, the data ratio for \cite{Eidelman:2004wy}
$f_K/f_\pi\simeq 1.22$ gives
\begin{eqnarray}
\frac{m_s}{\hat{m}} & \simeq & 2 \frac{f_K}{f_\pi} - 1 \; \simeq \; 1.44 \\ 
\Rightarrow \; m_s & \simeq & 1.44\; \hat{m} \; \simeq \; 486.0 \;\mbox{MeV}\;.
\label{eqmstrange1}  
\end{eqnarray}
Then the kaon GTR
\begin{equation}
f_K g \; = \; \frac{1}{2}(m_s+\hat{m}) \; = \; 411.75\; \mbox{MeV}
\label{eqgtrelkaon1}
\end{equation}
predicts, via $g=2\pi/\sqrt{3}$,
\begin{equation}
f_K \; = \; \frac{1}{2}(m_s+\hat{m})/(2\pi/\sqrt{3})\;=\;113.50\;\mbox{MeV}\;,
\end{equation}
giving the ratio
\begin{equation} \frac{f_K}{f_\pi}\; = \; \frac{113.50}{93} \; \simeq \; 1.22\; , 
\end{equation}
a second test of $g=2\pi/\sqrt{3}$ (Eq.~(\ref{eqgtr1}) being the first test).
Alternatively, from Eq.\ (\ref{eqmhat1}) with $f_\pi\simeq 93$~MeV,
$f_K/f_\pi\simeq 1.22$,
\begin{equation}
2 f_K g - \hat{m} \; = \; m_s \simeq 485.7\; \mbox{MeV}\; ,
\label{eqlast1}
\end{equation}
very close to Eq.~(\ref{eqmstrange1}), a third test of $g=2\pi/\sqrt{3}$.

ChPT appears not to alter these GTRs, nor does so PCAC, even though both were
first obtained theoretically via the L$\sigma$M (not with quarks, but nucleons
as fermions \cite{Gell-Mann:1960np}). Recall that ChPT rules out some kind of
L$\sigma$M in Appendix~B of Ref.~\cite{Gasser:1983yg}, which is an indirect
mark against ChPT, because our GTR-L$\sigma$M chiral scheme in
Eqs.~(\ref{eqgtr1})--(\ref{eqlast1}) above is a close match with data.
\section{Goldberger-Treiman discrepancies}
\label{sec5}
A \textit{once}-subtracted dispersion relation (containing \textit{no}
\/additional parameters) predicts (for $f_\pi = (92.42\pm 0.26)$~MeV
\cite{Eidelman:2004wy}) the $q^2$ variation of $f_\pi$ as \cite{Bugg:2003bj} 
\begin{equation}
\frac{f_\pi(m^2_\pi)}{f_\pi(0)} - 1 \; =\; \frac{m^2_\pi}{8\pi^2 f^2_\pi}
\left( 1 + \frac{m^2_\pi}{10 \hat{m}^2} \right) \; \simeq \; 2.84 \; \% \; .
\label{eqdisprel1}
\end{equation}
The dominant 2.79\% ${\cal O}(m^2_\pi)$ term in Eq.~(\ref{eqdisprel1}) was
first obtained in Refs.~\cite{Scadron:1981dn}. The smaller 0.05\%
${\cal O}(m^4_\pi)$ term was found in
Ref.~\cite{Nagy:2004tp}, having been anticipated numerically in 
Ref.~\cite{Bugg:2003bj}. In any case, Eq.\ (\ref{eqdisprel1}) is a Taylor
series in $m^2_\pi$, $m^4_\pi$, needing only \textit{two} \/terms.
Alternatively, we invoke data to explain the GT discrepancy as
\cite{Delbourgo:1998kg,Surovtsev:2002kr,Nagy:2004tp}
\begin{equation}
\Delta \; = \; 1 - \frac{m_N g_A}{f_\pi g_{\pi NN}} \; = \; (2.07 \pm 0.57)
\% \; ,
\label{eqdel1}
\end{equation}
using $m_N=938.92$~MeV, $|\,g_A/g_V|=1.2695\pm0.0029$ \cite{Eidelman:2004wy},
$f_\pi = (92.42\pm 0.26)$~MeV \cite{Eidelman:2004wy}, and $g_{\pi NN} =13.17
\pm0.06$ \cite{Bugg:2004cm}. Prior 1971 measurements gave $g_{\pi NN}\simeq
13.40$ and $\Delta\simeq 3.8\; \%$. The analysis of Sec.~\ref{sec2} yields
\begin{equation} \frac{\sigma_{\pi N}}{2 m_N} \; = \; \frac{63}{2\times 938.92}
\; \simeq \; 3.35\%\; .
\label{eqrat1}
\end{equation}
\textit{All} \/of the above analyses suggest a $3\%$ chiral-breaking effect.

However, the ChPT scalar form factors predict
\begin{equation} \frac{f_\pi(m^2_\pi)}{f_\pi(0)} - 1 \;\sim \;
\left\{ \begin{array}{rcl} 6.7\; \% & & \mbox{via one loop$\;\,$
\cite{Gasser:1983kx},} \\
7.2\; \% & & \mbox{via two loops \cite{Ananthanarayan:2004xy},} \\
8.2\; \% & & \mbox{via two loops \cite{Pelaez:2003eh}.} 
\end{array}\right.
\label{eqgtdiscrep1}
\end{equation}
Even though there is a two-loop debate as to how ChPT should proceed,
\textit{none} of these 6.7\%, 7.2\%, 8.2\% predictions
(based on scalar form factors) are near the 3\% chiral-breaking
model-independent predictions of Eqs.~(\ref{eqdisprel1}), (\ref{eqdel1}),
or (\ref{eqrat1}).
\section{Constituent quark mass via baryon magnetic moments}
\label{sec6}
The nonrelativistic valence (V) quark model has axial-vector spin components
of the nucleon \cite{Beg:1964nm}
\begin{equation}
\Delta u_V=\frac{4}{3} \;\; , \;\;\; \Delta d_V=-\frac{1}{3} \;\;\; , \;\;\;
\Delta s_V=0 \; ,
\label{axial}
\end{equation}
with total spin $\Sigma_V=\Delta u_V + \Delta d_V + \Delta s_V = 1$. Although
$\Delta u - \Delta d \approx 1.27$ \cite{Eidelman:2004wy} is about 30\% lower
than the valence value from Eq.~(\ref{axial}), good valence predictions from
the nucleon magnetic moments (m.m.) stem from \cite{Beg:1964nm}
\begin{eqnarray}
\mu_p(uud) \; = \; \mu_u\Delta u_V+\mu_d\Delta d_V+\mu_s\Delta s_V \; , \\
\label{mmp}
\mu_n(ddu) \; = \; \mu_d\Delta u_V+\mu_u\Delta d_V+\mu_s\Delta s_V \; ,
\label{mmn}
\end{eqnarray}
predicting the ratio
\begin{equation}
\frac{\mu_p}{\mu_n} \; = \; \frac{\left(\displaystyle\frac{\mu_u}{\mu_d}\,
\frac{\Delta u_V}{\Delta d_V}\right)+1+0}{\displaystyle\frac{\Delta u_V}
{\Delta d_V}+ \frac{\mu_u}{\mu_d}+0} \; = \; \frac{9}{-6} \; = \; -1.5 
\label{mmpnr}
\end{equation}
(using quark m.m.\ values $\mu_u=e/3m_u$, $\mu_d=-e/6m_d$), close to the
observed \cite{Eidelman:2004wy} ratio $2.793/(-1.913)\approx-1.46$.
Also, the nucleon m.m.\ difference \cite{Eidelman:2004wy,Beg:1964nm} is
\begin{equation}
\mu_p-\mu_n \; = \; (\mu_u-\mu_d)(\Delta u_V-\Delta d_V) \; = \; 
4.706\left(\!\frac{e}{2m_N}\!\right) \; ,
\label{mmpnd}
\end{equation}
predicting the average constituent quark mass (for $\hat{m}\approx m_u,m_d$)
\begin{equation}
\hat{m} \; = \; m_N \frac{5/3}{4.706} \; \approx \; 332.5 \; \mbox{MeV} \; ,
\label{mhatmm}
\end{equation}
for $\Delta u_V-\Delta d_V=5/3$, near $\hat{m}\approx m_p/\mu_p\approx 336$
MeV, and near the GT mass in Eq.~(\ref{eqgtr1}), or the anticipated m.m.\
quark mass in Eq.~(\ref{eqmhat1}).

As a matter of fact, assuming $\Delta s=0$, the sum $\mu_p+\mu_n$ predicts a
larger value $\hat{m}\approx355.6$ MeV, which reduces to
Eqs.~(\ref{eqgtr1},\ref{eqmhat1}) only if there is a slight strangeness
component in the nucleon, viz.\
\begin{equation}
\Delta s \; \approx \; \frac{337.5-355.6}{337.5} \; \approx \; -5.4\% \; .
\label{scn}
\end{equation}
This result is compatible with recent data \cite{Foudas:1989rk} finding
 $-\Delta s\leq6\%$.

In a similar manner, the $\Sigma$ to $N$ m.m.\ valence-difference ratio
is predicted as
\begin{equation}
\frac{\mu_{\Sigma^+}(uus)-\mu_{\Sigma^-}(dds)}{\mu_p(uud)-\mu_n(ddu)} = 
\frac{\Delta u_V-\Delta s_V}{\Delta u_V-\Delta d_V}  = \frac{4/3}{5/3} =
0.800 \;
\label{mmspn}
\end{equation}
only 4\% higher than the observed \cite{Eidelman:2004wy} difference ratio 
0.769. Moreover, the present $\beta$-decay value \cite{Eidelman:2004wy}
\begin{equation}
\Delta u - \Delta d \; = \; g_A \; = \; 1.2695\pm0.0029 \; ,
\label{udbeta}
\end{equation}
and the $\lambda_8$ component found from various hyperon semileptonic weak
decays give
\begin{equation}
\Delta u + \Delta d - 2\Delta s \; = \; g_A\left[\frac{3f-d}{f+d}\right]_A
\; \approx \; 0.584
\label{duds}
\end{equation}
(for the empirically determined ratio \cite{Dziembowski:1989sz}
$(d/f)_A\approx1.74$).
As we shall now see, Eq.~(\ref{duds}) requires a slight $\Delta s\neq0$,
i.e., $\Delta s=-5.7\%$.

Lastly, adding and subtracting Eqs.~(\ref{udbeta},\ref{duds}) leads to
\cite{Scadron:1997jq}
\begin{equation}
\Delta u - \Delta s \; \approx \; 0.927 \;\;\; , \;\;\;
\Delta d - \Delta s \; \approx \; -0.343 \; .
\label{dus}
\end{equation}
Then the latter four equations uniquely predict
\begin{equation}
\Delta u \approx 0.87 \; , \; \Delta d \approx -0.40 \; , \;
\Delta s \approx-0.057 \; , \; \Delta \Sigma \approx 0.41 \; ,
\label{dudss}
\end{equation}
together with the good valence results Eqs.~(\ref{mmpnr},\ref{mmpnd},\ref{scn})
and the constituent quarks mass Eq.~(\ref{mhatmm}). This quark-spin pattern
(\ref{dudss}) is compatible with the QCD predictions \cite{Ellis:1994py}
\begin{equation}
\begin{array}{l}
\Delta u\;=\;\;\;\,0.85\pm0.03 \;\;\; , \;\;\; \Delta d\;=\;-0.41\pm0.03 \; ,\\
\Delta s\;=\;-0.08\pm0.03 \;\;\; , \;\;\; \Delta\Sigma\;=\;\;\:0.37\pm0.07 \; .
\end{array}
\label{dudssqcd}
\end{equation}

Also, dynamical tadpole leakage is similar in spirit to the \qllsm, but now
with axial-vector $f_1$(1285), $f_1$(1420) mixing \cite{Anselmino:1991sy}.
This generates quark spins close to Eqs.~(\ref{dudss},\ref{dudssqcd}), with
$\Delta s=-6.0\%$, near the values $-5.7\%$ in Eq.~(\ref{duds}) and $-5.4\%$
in Eq.~(\ref{scn}).

Given this consistent pattern of quark spins with approximate average quark
mass in Eq.~(\ref{mhatmm}), for $\hat{m}=(m_u+m_d)/2$, $\mu_u=e/3m_u$,
$\mu_d=-e/6m_d$ and leading-order $\hat{m}=m_p/\mu_p\approx336.0$ MeV, and
further folding in $m_d-m_u\approx4$ MeV, we obtain (in MeV) a quadratic
equation for $\mhcon$, viz.\
\begin{equation}
\mhcon^2 - \frac{m_p}{2.792847}\,\left(\mhcon+\frac{14}{9}\right) \; = \; 0 \;,
\label{mcon}
\end{equation}
whose only positive solution is $\mhcon\approx337.5$ MeV, as anticipated in
Eq.~(\ref{eqmhat1}). The latter is near Eq.~(\ref{mhatmm}), nearer still to 
$m_p/\mu_p\approx336.0$ MeV, and even closer to the GTR quark mass in 
Eq.~(\ref{eqgtr1}). As mentioned in the Introduction, this bulk constituent
quark mass can be decomposed into its CL dynamically generated part $\mdyn$
and its chiral-broken current-quark-mass component $\mcur$:
\begin{equation}
\mcon \; = \; \mdyn + \mcur \; .
\label{mcodycu}
\end{equation}
Recall that $\mdyn$ was first estimated \cite{Delbourgo:1982tv}, via
quark-dressing and binding equations, as $\mdyn\sim$ 300--320 MeV, or
ideally
\begin{equation}
\mdyn \; = \; \frac{m_N}{3} \; \approx \; 313\;\mbox{MeV} \, .
\label{mdynnuc}
\end{equation}
This scale matches infrared QCD
\cite{Elias:1984aw,Delbourgo:1982tv,Bass:1994jx}, for $\alpha_s\approx0.5$
at a 1 GeV cutoff,
\begin{equation}
\mdyn \; = \; \left[\frac{4\pi}{3}\alpha_s\left<-\bar{q}q\right>\right]^
{\frac{1}{3}} \; \approx \; 313 \; \mbox{MeV} \; ,
\label{mdynqcd}
\end{equation}
for $\left<-\bar{q}q\right>\approx(245$ MeV$)^3$. Also, from Sec.~III, recall
that 
\begin{equation}
\mdyn \; = \; r^{-1}_\pi \; \approx \; 313 \; \mbox{MeV} \; ,
\label{mdynrpi}
\end{equation}
for VMD pion charge radius $r_\pi\approx0.63$ fm, near data
\cite{Eidelman:2004wy}.

As for the ChSB current-quark-mass scale, the ChPT review of 1982
\cite{Gasser:1982ap} takes
\begin{equation}
\mhcur \; \approx \; \frac{(f_\pi m_\pi)^2}{2\left<-\bar{q}q\right>} \;
\approx \; 5.5 \; \mbox{MeV} \; ,
\label{mhcur}
\end{equation}
for $f_\pi\approx93$ MeV and $\left<-\bar{q}q\right>\approx(245$ MeV$)^3$.
Reference \cite{Gasser:1982ap} implies this relation is due to GMOR
\cite{Gell-Mann:1968rz}, but Eq.~(\ref{mhcur}) and the $f_\pi$ scale are
hard to find in GMOR, which focuses instead on the good-bad structure of
$\bar{q}\lambda_i q$ and $\bar{q}\gamma_5\lambda_i q$ matrix elements
\cite{Fuchs:1979iw}. In fact, Ref.~\cite{Gunion:1976nt} stresses on page 462
that this GMOR $SU(3)$ assumption (leading directly to our Eq.~(\ref{mhcur}))
may not be correct. More recently, Ref.~\cite{Bass:1994jx} noted that,
if Eq.~(\ref{mhcur}) were true and if $\left<N|\bar{s}s|N\right>=0$ (in fact,
$\Delta s\approx-0.057$ is compatible with data \cite{Foudas:1989rk}), then it
is easy to show that the $\pi N$ $\sigma$ term should be in theory
\begin{equation}
\sigma_{\pi N}^{\mbox{\scriptsize th}} \; = \; \frac{3(m_\Xi-m_\Lambda)}
{\displaystyle\left(\frac{m_s}{\hat{m}}\right)_{\scur}-1} \; \approx \;
\frac{606\;\mbox{MeV}}{\displaystyle\left(\frac{m_s}{\hat{m}}\right)_{\scur}-1}
\; .
\label{pinsigth}
\end{equation}
If we follow recent ChPT with \cite{Gasser:1990ce} $\sigma_{\pi N}^{\chpt}=45$
MeV, then the latter equation requires $(m_s/\hat{m})_{\scur}\approx14.47$,
and so $\mscur\approx80$ MeV for $\mhcur\approx5.5$ MeV. \em ``If this were
true then a significant fraction of the nucleon's mass would be due to
strange quarks --- in contradiction with the quark model'' \em
\cite{Bass:1994jx}. Moreover, such a result would be in conflict with the
usual ChPT ratio \cite{Gasser:1982ap} $(m_s/\hat{m})_{\scur}\approx25$. We
shall return to current quarks in the next section.

\section{Effective current quark and Chiral-Symmetry-Breaking pion, kaon,
\bom{\eta_8} masses}
\label{sec7}
Following Ref.~\cite{McKellar:1986dr}, we use QCD to express a
momentum-dependent dynamical quark mass as 
\begin{equation} \mdyn(p^2) \; = \; \frac{\mdyn^3}{p^2} \; , 
\end{equation}
so that $\mdyn$ at $p^2=\mdyn^2$ is, as required by consistency,
\begin{equation} \mdyn (p^2 = \mdyn^2) \; \equiv \; \mdyn \; .
\end{equation}
Furthermore, we take the nonstrange constituent quark mass $\mhcon$ as
337.5 MeV from Secs.~\ref{sec4},\ref{sec6} and compute the effective
current quark mass as (avoiding any good-bad assumptions besides
Eq.~(\ref{mhcur}), but keeping the decomposition Eq.~(\ref{mcodycu}), and also
Eqs.~(\ref{mdynnuc}--\ref{mdynrpi}))
\begin{equation} \delta \hat{m} \; = \; \mhcon -
\frac{\mdyn^3}{\mhcon^2} \; \simeq \; 68.3 \; \mbox{MeV} \; , 
\label{dmhat}
\end{equation}
with $\delta \hat{m}\rightarrow 0$ when $\mhcon\rightarrow\mdyn$ in the CL,
as one expects for the current quark mass. Note that $\delta\hat{m}\simeq 68.3
\; \mbox{MeV}$ is very near the chiral-breaking $\sigma_{\pi N}\simeq
63$--$66$~MeV of Sec.~\ref{sec2}, also as expected!

Then for the $\bar{q}q$ pion, its mass is predicted from Eq.~(\ref{dmhat}) as
\begin{equation}
m_\pi \; = \; \delta \hat{m} + \delta \hat{m} \; \simeq \;
136.6 \; \mbox{MeV} \; ,
\label{eqmpi1}
\end{equation}
midway between the observed \cite{Eidelman:2004wy} $m_{\pi^0}=134.98$~MeV
and $m_{\pi^\pm}=139.57$~MeV.

Instead, the ChPT prediction~(\ref{mhcur}) suggesting $\mucur\sim5$ MeV and
$\mdcur\sim9$ MeV, or taking the ChPT scheme of Ref.~\cite{Colangelo:2003hf}
having finite-volume effects, do \textit{not} \/predict a pion mass anywhere
in the region of 140~MeV.
The QLL$\sigma$M assumes standard pion $\bar{q}q$ states $\left|\pi^+\right>=
\left|\bar{d}u\right>$, $\left|\pi^-\right>=\left|\bar{u}d\right>$,
$\left|\pi^0\right>=\left(\left|\bar{u}u\right>-
\left|\bar{d}d\right>\right)/\sqrt{2}$, and takes the pion as a tightly bound
$\left|\bar{q}q\right>$ Nambu-Goldstone ``fused'' nonstrange meson
(verified via the QLL$\sigma$M, VMD, and measured meson charge radius
in Sec.~\ref{sec3}). The observed proton charge radius \cite{Eidelman:2004wy}
$R_p=(0.870\pm 0.008)$~fm suggests the proton is a ``touching'' quark-pyramid
$uud$ state \cite{Scadron:2003yg} and there is minimal
$\left<\bar{s}s\right>_N$, either from data \cite{Foudas:1989rk}, phenomenology
\cite{Anselmino:1989br}, or from the magnetic-moment scheme of Sec.~\ref{sec6}.

Instead, we return to studying the nonstrange quark mass $\mhcur$, recalling
that $\mhcur = \delta\hat{m} \approx 68.3$~MeV in Eq.~(\ref{dmhat}) (also see
Ref.~\cite{Elias:1988ru}) successfully predicts the pion mass at
$m_\pi = 2\,\delta \hat{m} = 136.6$~MeV in Eq.~(\ref{eqmpi1}). 

Moreover, $\hat{m}_{cur}$ is sometimes called the current mass via neutral PCAC
(for a review, see Ref.~\cite{Scadron:1981dn}). Using quark structure
functions, one predicts the pion mass (squared) as
\begin{eqnarray}
m^2_\pi & = & 2\,\hat{m}^2_{cur}\,\bar{h}\;\;\;,\;\;\;\bar{h}\;=\;
\frac{5}{2}  \nonumber \\
\Rightarrow & & \hat{m}_{cur} \; = \; \frac{m_\pi}{\sqrt{5}} \; \approx \; 62.4
\; \mbox{MeV} \; ,
\label{eqourchsb1}
\end{eqnarray} 
for $m_{\pi^\pm}\approx 139.57$~MeV, invoking the spectator-helicity rule
\cite{Fuchs:1979iw}. Note from Eqs.~(\ref{eqourchsb1}) that here
$m^2_\pi \propto \hat{m}^2_{cur}$, whereas GMOR \cite{Gell-Mann:1968rz} and
ChPT take $m^2_\pi \propto \hat{m}_{cur}$. Fubini and Furlan \cite{Fubini:1965}
suggested that the r.h.s.\ of Eqs.~(\ref{eqourchsb1}) can behave as either
$\hat{m}_{cur}$ or $\hat{m}^2_{cur}$, depending on the renormalization scale.
At a 1~GeV scale we can write
\begin{equation}
m^2_\pi \; = \; \left<\pi\right|H_{\chsb}\left|\pi\right>\;
\sim \; \hat{m}_{cur} \, \left<\pi\right|\bar{u}u+\bar{d}d\left|\pi\right> \; ,
\label{mpis}
\end{equation}
or roughly, from Eq.~(\ref{mpis}) given Eq.~(\ref{mcodycu}) and
$\mdyn\approx313$ MeV,
\begin{eqnarray}
m^2_\pi & \sim &  \mhcur \, (\mdyn + \mhcur) \;\; \Rightarrow
\;\; \mhcur \; \sim \; 53\; \mbox{MeV} \, , \nonumber \\
\label{eqourchsb2}
\end{eqnarray}
taking $\mdyn \simeq 313$~MeV as in Sec.~\ref{sec6}.

Alternatively, we can take the baryon $d/f$ ratio or structure-function
integrals as \cite{Fuchs:1979iw}
\begin{eqnarray}
\frac{d}{f} & = & \frac{\bar{f}_u-2\bar{f}_d+\bar{f}_s}
{\bar{f}_u-\bar{f}_s} \nonumber \\
& = & - \frac{3}{5} \, \frac{3\, m^2_\Sigma - m^2_N - m^2_\Lambda
- m^2_\Xi}{m^2_\Xi-m^2_N} \; \approx \; - 0.28 \; , \nonumber \\
\frac{\bar{f}_d}{\bar{f}_u} & \approx & 0.64 \;\;\; , \;\;\;
\bar{f}_s \; = \; 0 \;\;\; \mbox{(near $\bar{f}_s \; \approx \; -0.057$)} \; .
\label{fh}
\end{eqnarray}
In fact, the spectator-helicity rule \cite{Fuchs:1979iw} is slightly altered to
(using squared baryon masss as in Eq.~(\ref{fh}), rather than the
$\bar{f}_u=7.9$, \ldots\ as found in Eq.~39 of Ref.~\cite{Fuchs:1979iw})
\begin{eqnarray}
\bar{h} & = & \frac{5}{2} \; , \quad \bar{f}_u \; \approx \;
7.64 \; , \quad \bar{f}_d \; \approx \; 4.85 \; , \nonumber \\
\frac{\bar{f}_d}{\bar{f}_u} & \approx & 0.635 \;\; , \;\;
\bar{f}_u + \bar{f}_d \; \approx \; 12.49 \;\; \mbox{for} \;\;
\bar{f}_s \; = \; 0 \; .
\label{shr}
\end{eqnarray}
Note that the latter ratio is compatible with $\bar{f}_d/\bar{f}_u=0.64$
in Eq.~(\ref{fh}), and the latter sum implies the Jaffe--Llewellyn-Smith form
\cite{Jaffe:1972yv} for the current quark mass (squared) is
\begin{eqnarray} \mhcur^2 & = & \frac{m_N\,\sigma_{\pi N}}{\bar{f}_u +
\bar{f}_d} \nonumber \\[1mm]
& = & \frac{938.9 \; \mbox{MeV} \; \times \; 63\;\mbox{MeV}}{12.49}
\; = \; (68.8\;\mbox{MeV})^2 \, , \;\;\;  \label{eqourchsb3}
\end{eqnarray}
very near the predicted effective current quark mass of 68.3 MeV in 
Eq.~(\ref{dmhat}).
Thus, we deduce in several independent ways (Eqs.~(\ref{eqourchsb1}),
(\ref{eqourchsb2}), (\ref{eqourchsb3})) that $\hat{m}_{cur} \simeq 62.4$~MeV,
$\sim53$~MeV, $\sim68.8$~MeV, all near $\delta\hat{m} \approx 68.3$~MeV from
Eq.~(\ref{dmhat}), but \textit{not} \/near the ChPT value
$\mhcur\approx5.5$ MeV in Eq.~(\ref{mhcur}).

Also, following Ref.~\cite{Gunion:1976nt}, we have
\begin{equation}
\begin{array}{l}
m^2_\Sigma-m^2_N\;=\;\bar{f}_d\,(m^2_s-\hat{m}^2)_{\mbox{\scriptsize cur}}\;=\;
0.5417\;\mbox{GeV}^2 \; , \\[1mm]
m^2_\Xi-m^2_N \; = \;\bar{f}_u\,(m^2_s-\hat{m}^2)_{\mbox{\scriptsize cur}}\;=\;
0.8556\;\mbox{GeV}^2 \; .
\end{array}
\label{msksn}
\end{equation}
This leads to the ratio $\bar{f}_d/\bar{f}_u\approx0.633$ (independent of the
$(m_s^2\!-\!\hat{m}^2)_{\scur}$ scale), near the value 0.64 found
above. Moreover, if we sum the two equations in Eq.~(\ref{msksn}) and use
$\bar{f}_u+\bar{f}_d\approx12.49$ from Eq.~(\ref{shr}), we get
$(m_s^2\!-\!\hat{m}^2)_{\scur}\approx(1.3973/12.49)$ GeV$^2\approx0.1119$
GeV$^2$. Substituting then the value $\mhcur=0.0683$ GeV from Eq.~(\ref{dmhat})
yields the consistent ratios
\begin{equation}
\left(\frac{m_s}{\hat{m}}\right)_{\!\mbox{\scriptsize cur}} \; \approx \;
\sqrt{\frac{0.1119}{(0.0683)^2}+1} \; = \; 5.00 
\label{msmhat}
\end{equation}
from $(\bar{f}_u+\bar{f}_d)$ above,
\begin{equation}
\left(\frac{m_s}{\hat{m}}\right)_{\!\mbox{\scriptsize cur}} \; \approx \;
\sqrt{\frac{0.1125}{(0.0683)^2}+1} \; = \; 5.01
\label{msmmhat}
\end{equation}
from $(\bar{f}_u-\bar{f}_d)$ above,
\begin{equation}
\left(\frac{m_s}{\hat{m}}\right)_{\!\mbox{\scriptsize cur}} \; = \;
\sqrt{\frac{2m^2_K}{m^2_\pi}-1} \; = \; 5.00 
\label{chikaon}
\end{equation}
from the PCAC $K$-to-$\pi$ ratio \cite{Elias:1988ru}, for $m_K^2=13m_\pi^2$,
\begin{equation}
\left(\frac{m_s}{\hat{m}}\right)_{\!\mbox{\scriptsize cur}} \; = \;
\frac{2m_K}{m_\pi}-1 \; = \; 6.21 
\label{lcone}
\end{equation}
from the light cone \cite{Sazdjian:1974gk}, and
\begin{equation}
\left(\frac{m_s}{\hat{m}}\right)_{\!\mbox{\scriptsize cur}} \; = \;
\sqrt{\frac{3\Delta m^2_N}{m_N\sigma_{\pi N}}+1} \; = \; 4.93
\label{dmps}
\end{equation}
from \cite{Elias:1988ru} $\Delta m^2_N=0.46$ GeV$^2$, being the
nucleon mass shift (squared) from the baryon-octet $SU(3)$ value of
(1.158 GeV)$^2$. The average ratio from Eqs.~(\ref{msmhat}--\ref{dmps})
is 5.23, which when combined with $\mhcur=68.3$ MeV implies
$\mscur=357$~MeV, very near the analogue of
Eq.~(\ref{dmhat}), i.e., $\mscur=\mscon-\mdyn^3/\mscon^2=356$ MeV,
for $\mscon=486$ MeV and $\mdyn=313$ MeV.
Note that the ChPT ratio $(m_s/\hat{m})_{\scur}\approx25$ scaled to
$\mhcur\approx5.5$ MeV predicts $\mscur\approx137.5$ MeV, leading to an
\textit{inconsistent} \/picture of Nature.

Finally, we extend the $\pi\pi$ $\sigma$ term
$\sigma_{\pi\pi}\approx m_\pi^2$ from Eq.~(\ref{eqsigpipi1}) to the three
ChSB meson $\sigma$ terms
\begin{equation}
\sigma_{\pi\pi} \;\;\; , \;\;\; \sigma_{KK} \;\;\; , \;\;\;
\sigma_{\eta_8\eta_8} = m^2_\pi\left(1,\frac{1}{2},\frac{1}{3}\right)
\label{sigmaterms}
\end{equation}
as found in Refs.~\cite{Ling-Fong:1971eg}. Note that all three vanish in the
CL, with $m^2_\pi\to0$. To obtain the chiral-broken $\eta_8$ mass, we invoke
the Gell-Mann--Okubo (GMO) value (see also Eqs.\ (\ref{modindep4}))
\begin{equation}
m^{\gmo}_{\eta_8} \; = \; \sqrt{\frac{4m^2_K-m^2_\pi}{3}} \; \approx \;
\sqrt{17}m_\pi \; \approx \; 563\;\mbox{MeV} \; ,
\label{etaeight}
\end{equation}
using $m^2_K\approx13m^2_\pi$, and the average ChSB pion mass 136.6 MeV from
Eq.~(\ref{eqmpi1}). Since the observed \cite{Eidelman:2004wy}
$m_\eta$ is 547.75 MeV, $m^{\gmo}_{\eta_8}\approx563$ MeV from Eq.~(\ref{etaeight})
is about 2.8\% greater than $m_\eta$, similar to the 3\% GTR discrepancies in
Sec.~V, but much lower than the 6.7--8.2\% ChPT predictions in
Eq.~(\ref{eqgtdiscrep1}). Moreoever, folding the GMO relation into the three
meson $\sigma$ terms of Eq.~(\ref{sigmaterms}), one finds
$3\sigma_{\eta_8\eta_8}\!-4\sigma_{KK}\!+\sigma_{\pi\pi}=0$, as
anticipated in Eqs.~(\ref{sigmaterms}) and (\ref{etaeight}), giving a
consistent pattern $(3/3)-(4/2)+1=0$.

\section{Ground-state scalar \lowercase{\bom{\bar{q}q}} nonet}
\label{sec8}
It is well-known and commonly accepted that there exists (within a $U(3)\times
U(3)$ flavor scheme) a nonet of pseudoscalar mesons ($\pi(137)$, 
$K(496)$, $\eta(548)$, $\eta^\prime(958)$) that play the role of 
Goldstone bosons associated with ChSB. What is not so well-known and 
accepted is that there is also experimental evidence \cite{Eidelman:2004wy}
for a corresponding light scalar-meson nonet
\cite{Scadron:1982eg,VanBeveren:1986ea} $f_0(600)$ ($\sigma$), $K_0^*(800)$
($\kappa$), $f_0(980)$, $a_0(985)$. The members of the latter nonet, being
much too light
to be accomodated as naive (unquenched) quark-model states, are rather to be 
interpreted as the chiral partners (see e.g.\ Ref.~\cite{vanBeveren:2002mc})
of the former pseudoscalar mesons. In this 
spirit, it is useful to define the following field matrices $S(x)$ and 
$P(x)$, for $U(3)\times U(3)$  flavor nonets of scalar and pseudoscalar 
mesons, respectively:
\begin{equation} S = \left(
\begin{array}{ccc} \sigma_{u\bar{u}} & a^+_0 & \kappa^+ \\[1mm]
a^-_0 & \sigma_{d\bar{d}} & \kappa^0 \\[1mm]
\kappa^- & \bar{\kappa}^0 & \sigma_{s\bar{s}}
\end{array}\right) \;\;\; , \;\;\; P = \left(
\begin{array}{ccc} \eta_{u\bar{u}} & \pi^+ & K^+ \\[1mm]
\pi^- & \eta_{d\bar{d}} & K^0 \\[1mm]
K^- & \bar{K}^0 & \eta_{s\bar{s}}
\end{array}\right) \; .
\label{spmat}
\end{equation}
For later convenience, we also define $\sigma_{n} \equiv (\sigma_{u\bar{u}}
+ \sigma_{d\bar{d}})/\sqrt{2}$, $\sigma_{3}\equiv (\sigma_{u\bar{u}} - 
\sigma_{d\bar{d}})/\sqrt{2}\simeq a^0_0$,
  $\eta_{n} \equiv (\eta_{u\bar{u}} + \eta_{d\bar{d}})/\sqrt{2}$, and 
$\eta_{3}\equiv (\eta_{u\bar{u}} - \eta_{d\bar{d}})/\sqrt{2}\simeq \pi^0$.

The still somewhat controversial status of the lightest members of the
ground-state scalar-meson nonet, namely the $\sigma$(600) and the
$\kappa$(800), is due to both experimental and theoretical difficulties.
On the one hand, the amplitudes in elastic $\pi\pi$ and $K\pi$ scattering
rise very slowly from threshold upwards, due to nearby Adler zeros
\cite{Bugg:2003kj,Rupp:2004rf} just below threshold. On the other hand, the
theoretical description of light scalar-meson resonances turns out to be quite
cumbersome, owing to the large unitarization effects, which demand a manifestly
nonperturbative treatment.

In a field-theoretic framework, the most efficient and obvious approach to the
light scalars is indubitably the \qllsm\
\cite{Delbourgo:1993dk,Scadron:2003yg,Delbourgo:1998kg,Scadron:2002mm},
which contains from the outset --- contrary to e.g.\ QCD --- all the relevant
(experimentally observable) degrees of freedom, i.e., quarks as well as
scalar and pseudoscalar mesons. Moreover, already at tree level 
important nonperturbative features of strong interactions are included. The 
interaction Lagrangian of (anti)quarks and (pseudo)scalar mesons in the 
QLL$\sigma$M is given by ${\cal L}_{int} (x) = \sqrt{2}\,g\,\bar{q}(x) 
(S(x) + i P(x) \,\gamma_5) q(x)$, with $g$ the strong coupling
constant given by $|g|\simeq 2\pi/\sqrt{3}$, as follows from
one-loop dynamical generation \cite{Delbourgo:1993dk,Kleefeld:2005hd}. By 
integrating out (anti)quarks, the QLL$\sigma$M Lagrangian can be converted 
into an effective $U(3)\times U(3)$ L$\sigma$M meson Lagrangian 
displaying strong similarity with the well-known ``traditional'' $U(3)\times 
U(3)$ L$\sigma$M Lagrangian \cite{Levy:1967a,Gasiorowicz:1969kn,
Cabibbo:1970uc,Delbourgo:1998kg,Kleefeld:2005qs}, briefly summarized in
Appendix~\ref{appendix1}.

The QLL$\sigma$M (without (axial-)vector mesons) in the CL ($m^2_\pi=m_K^2=0$)
predicts \cite{Delbourgo:1982tv,Scadron:1982eg,Delbourgo:1998kg} the
following NJL-like \cite{Nambu:1961tp} mass relations and chiral-limiting
scalar-meson masses:
\begin{eqnarray}  (m^{\cl}_{\sigma_n})^2 & = & \makebox[2.6cm][l]{$(2 
\hat{m}_{\cl})^2$}\; \simeq \; (626\; \mbox{MeV})^2 \; ,
\label{eqmesmasscl2} \\
(m^{\cl}_{\sigma_s})^2 & = & \makebox[2.6cm][l]{$(2\,m^{\cl}_s)^2$}\; 
\simeq \; (901\; \mbox{MeV})^2 \; ,
\label{eqmesmasscl3} \\
(m^{\cl}_{\kappa})^2 & = & \makebox[2.6cm][l]{$(2\hat{m}_{\cl}) (2 
m^{\cl}_s)$} \; \simeq \; (751\; \mbox{MeV})^2 \; ,
\label{eqmesmasscl4}
\end{eqnarray}
with $\hat{m}_{\cl}=313$~MeV, $m^{\cl}_s=1.44\times 
\hat{m}_{\cl}=450.72$~MeV.
Beyond the CL, the leading contributions to the scalar-meson masses due to
quark loops will lead to expressions equivalent to
Eqs.~(\ref{eqmesmasscl2}--\ref{eqmesmasscl4}), but now with
non-CL quark masses, i.e. (see e.g.\ Ref.~\cite{Scadron:2003yg}),
\begin{eqnarray} m_{\sigma_n}^2 & \simeq & \makebox[1.6cm][l]{$(2 
\hat{m})^2$}\; = \; (675\; \mbox{MeV})^2 \; ,
\label{eqmesmassx2} \\
m_{\sigma_s}^2 & \simeq & \makebox[1.6cm][l]{$(2 m_s)^2$}\; = \; (972\; 
\mbox{MeV})^2 \; ,
\label{eqmesmassx3} \\
m_{\kappa}^2 & \simeq & \makebox[1.6cm][l]{$(2 \hat{m})(2m_s)$} \; = \; 
(810\; \mbox{MeV})^2 \; ,
\label{eqmesmassx4}
\end{eqnarray}
with $\hat{m}=337.5$~MeV, $m_s=1.44\times \hat{m}=486.0$~MeV, which
predictions lie impressively close to the experimental values.\footnote{The PDG \cite{Eidelman:2004wy} further states that the nonstrange 
process $a_1(1260)\rightarrow \sigma \pi$ is the dominant mode, and that 
$a_1(1260)\rightarrow f_0(980) \pi$ is \textit{not} \/seen, presumably 
because the $f_0(980)$ is (approximately) a scalar $\bar{s}s$ state 
\cite{vanBeveren:2000ag,Kleefeld:2001ds}.}

A realistic nonstrange-strange mixing angle of $\phi_s\simeq \pm 18^\circ$ 
\cite{Kleefeld:2001ds} implies, with the help of
Eqs.~(\ref{eqsigmix1a}) and (\ref{eqsigmix2a}), the following values for 
$m_\sigma$ and $m_{f_0}$:
\begin{eqnarray} m^2_\sigma & = & \frac{1}{2} \left( 
m^2_{\sigma_s}+m^2_{\sigma_n} - 
\frac{m^2_{\sigma_s}-m^2_{\sigma_n}}{\cos(2\phi_S)}\right) \nonumber \\
 & = & (630.8\; \mbox{MeV})^2 \; ,
\label{eqsigmix1} \\[2mm]
m^2_{f_0} & = & \frac{1}{2} \left( m^2_{\sigma_s}+m^2_{\sigma_n} + 
\frac{m^2_{\sigma_s}-m^2_{\sigma_n}}{\cos(2\phi_S)}\right) \nonumber \\
  & = & (1001.3\; \mbox{MeV})^2 \; .
\label{eqsigmix2} 
\end{eqnarray}
Of course, the latter mass predictions are
still subject, in principle, to further corrections stemming from
non-zero pseudoscalar-meson masses, as can be seen from
Eqs.~(\ref{eqmesmassnew1}--\ref{eqmesmassnew4}) predicted by the
``traditional'' \lsm\ approach discussed in the Appendix. However,
as noticed there, several uncertainties persist within such a
formalism, which make it hard to establish a one-to-one correspondence
with the more straightforward \qllsm.

Comparing now with the experimental scalar-meson masses, we see that the
non-CL \qllsm\ results do a very good job. Starting with the $\sigma$ meson,
the PDG tables \cite{Eidelman:2004wy} very cautiously list the $f_0$(600) with
an extremely wide mass range of 400--1200~MeV, but giving a large number of 
recent experimental findings or analyses tending to converge around a typical
mass value of about 600 MeV. Also theoretical coupled-channel models reproducing
the corresponding $S$-wave $\pi\pi$ phase shifts and accounting for a low-lying
$\sigma$ pole lead to similar predictions.
For instance, the unitarized model of Ref.~\cite{VanBeveren:1986ea} found
a pole at $(470-i\times208)$ MeV, with a peak mass around 600 MeV.
More recently, the $\pi\pi\rightarrow \pi\pi,K\bar{K}$ analysis 
in Ref.~\cite{Surovtsev:2002kr} found an even broader $\sigma$
pole, with a real part of 570--600 MeV, referring to the $\sigma$ as the
``$f_0(665)$''. Although the precise \textit{mass} \/to be attributed
to a broad resonance is always model-dependent, the latter results are 
fully compatible with the value $2\hat{m}\simeq 675$~MeV of Eq.\ 
(\ref{eqmesmassx2}), or the more realistic value $630.8$~MeV of Eq.\ 
(\ref{eqsigmix1}).

As for the $I=1/2$ scalar $\kappa$ meson, seemingly doomed not so long ago
\cite{Cherry:2000ut}, despite our old theoretical predictions
\cite{Scadron:1982eg,VanBeveren:1986ea}, it was first experimentally
rehabilitated by the E791 collaboration \cite{Aitala:2002kr}, providing a
mass value of
\begin{equation}
m_{\kappa} \; = \; (797\pm 19)\; \mbox{MeV} \; .
\label{E791}
\end{equation}
This is to be compared to 810 MeV from Eq.~(\ref{eqmesmassx4}), 800~MeV from
Refs.~\cite{Scadron:1982eg,Paver:1983ys}, as well as the pole positions
$(727-i\times263)$ MeV \cite{VanBeveren:1986ea} and
$(714-i\times228)$ MeV \cite{vanBeveren:2001kf}. Note that the latter two
coupled-channel model calculations both give a peak cross section around
800 MeV. On the other hand, a very recent analysis \cite{Bugg:2005xz} of
$K\pi$ data from different high-statistics experiments yields the pole position
\begin{equation}
m_{\kappa} \; = \; (760\pm20\pm40)-i\times(420\pm45\pm60) \; \mbox{MeV} \; .
\label{Bugg05}
\end{equation}
As a final remark on the $\kappa$, we should stress the importance of 
unitarization \cite{Kleefeld:2005qs,Kleefeld:2005xx} effects, which are capable
of e.g.\ shifting the Lagrangian mass parameter in Eq.~(\ref{eqkappamass1})
from 1128 MeV to a pole value of $\sim$700--800 MeV, besides dynamically
generating extra states
\cite{VanBeveren:1986ea,vanBeveren:2001kf,Tornqvist:1995ay}. However, in
the latter reference, the authors introduce by hand a rather unphysical Adler
zero, which apparently moves the dynamically generated $\kappa$ pole far away
from the physical region \cite{Rupp:2004rf}.

Next we discuss the $f_0$(980). The mass of $(980\pm10)$ MeV listed by the PDG
\cite{Eidelman:2004wy} is entirely compatible with the above values of 972 MeV
and 1001.3 in Eqs.~(\ref{eqmesmassx3}) and (\ref{eqsigmix2}), respectively. So
are the $f_0$(980) pole positions of $\{(998\pm4)-i\times(17\pm4)\}$ MeV and
$(994-i\times14)$ MeV given in the very recent analysis of
Ref.~\cite{Bugg:2005xz}.

Concerning the $a_0$(985), the coupled-channel approach of
Ref.~\cite{VanBeveren:1986ea}, which includes the crucial $\eta\pi$ and
$K\bar{K}$ channels, predicts a reasonable pole mass of $(968-i\times28)$ MeV,
to be compared to $(1036-i\times84)$ MeV in the experimental analysis in
Ref.~\cite{Bugg:2005xz}. As already noted in 
Ref.~\cite{Scadron:2003yg}, we consider the approximate mass degeneracy of the
$a_0$(985) and $f_0$(980) purely accidental, even though the theoretical
situation in a \qllsm/\lsm\ framework is still unsettled, due to the apparent
relevance of meson loops \cite{Scadron:2003yg}. Namely, were we to neglect
meson contributions to the $a_0$(985) mass in Eq.~(\ref{eqmesmassnew1}), then
it would be approximately degenerate with the $\sigma_n$. On the other hand, if
we simply evaluate Eq.~(\ref{eqmesmassnew1}) with $\hat{m}=337.5$ MeV, $\xi=1$,
and $\phi_P =(41.2\pm 1.1)^\circ$ \cite{Kleefeld:2005qs,Kroll:2005sd}, we
obtain a surprisingly accurate mass prediction of 1012 MeV. The more realistic
coupling ratio $\xi=0.9064$ (see our discussion in the Appendix) would even
imply $m_{a_0}=990$~MeV.

An alternative, more empirical way to estimate scalar-meson masses is by
using quadratic mass differences motivated by the IMF. The resulting
equal-splitting laws have a remarkable predictive power (see e.g.\
Ref.~\cite{Scadron:2003yg}, Appendix C),
being fully compatible with the foregoing \qllsm\ predictions.

While these $\sigma_n(675)$, $\kappa(810)$, $\sigma_s(972)$, $a_0(985)$
$\bar{q}q$ scalars are in agreement, to a very good approximation, with the
QLL$\sigma$M theory, even without unitarization, one-loop (non-unitarized)
ChPT is \textit{not} \/compatible with a specific L$\sigma$M, as shown in
Appendix B of Ref.~\cite{Gasser:1983yg}. Therefore, non-unitarized ChPT is at
odds with the entire ground-state scalar $\bar{q}q$ nonet\footnote
{Very recently \cite{Caprini:2005zr},  Roy equations have been used to extract
a $\sigma$ pole from low-energy $\pi\pi$ scattering data, in the framework of
ChPT. However, the given pole position, especially the claimed very small error
on it, is highly questionable in view of the neglect of the $K\bar{K}$
channel.},
as summarized in this section, if indeed the L$\sigma$M of
Ref.~\cite{Gasser:1983yg} possesses similar properties as the QLL$\sigma$M
discussed above.

\section{\bom{I=0} \bom{\pi\pi} scattering length}
\label{sec9}
Using the L$\sigma$M, PCAC, and crossing symmetry, Weinberg originally
predicted \cite{Weinberg:1966kf} for the $I=0$, $J=0$ $\pi\pi$ scattering
length
\begin{equation} a_{00} \; = \; \frac{7}{32\pi} \frac{m_\pi}{f^2_\pi} \;
\simeq \; 0.159\; m^{-1}_{\pi}\; ,
\label{eqweinsc1}
\end{equation}
for $m_{\pi^+} = 139.57$~MeV, $f_\pi \simeq 92.42$~MeV \cite{Eidelman:2004wy}.

Stated slightly differently, and following V.~de Alfaro {\em et al.} \/(see
Ref.~\cite{Gell-Mann:1960np}), when working at the soft point $s=m^2_\pi$, one
sees that the net $\pi\pi$ amplitude ``miraculously vanishes'' (via chiral
symmetry), i.e.,
\begin{eqnarray}
M^{\mbox{\scriptsize net}}_{\pi\pi}&=&M^{\mbox{\scriptsize contact}}_{\pi\pi}+
M^{\mbox{\scriptsize $\sigma$-pole}}_{\pi\pi} \nonumber \\
 & = & \lambda + 2 g^2_{\sigma\pi\pi} (m^2_\pi - m^2_\sigma) \; = \; 0 \; ,
\label{eqampsum1}
\end{eqnarray}
since $g_{\sigma\pi\pi} = (m^2_\sigma-m^2_\pi)/(2\,f_\pi) = \lambda f_\pi$
from the L$\sigma$M Lagrangian. Thus, the contact term ``chirally eats'' the
$\sigma$ pole at $s=m^2_\pi$, yielding the famous amplitude ``Adler zero''
slightly below the $\pi\pi$ threshold \cite{Bugg:2003kj}. Crossing symmetry
then extends Eq.~(\ref{eqampsum1}) to \cite{Ko:1994en}
\begin{eqnarray}
M^{abcd}_{\pi\pi} (s,t, u) & = & A(s,t,u) \, \delta^{ab}\delta^{cd} \nonumber\\
 & + & B(s,t,u) \, \delta^{ac}\delta^{bd} \nonumber \\
 & + & C(s,t,u) \, \delta^{ad}\delta^{bc}  \; ,
\end{eqnarray}
with
\begin{eqnarray}
A^{\mbox{\scriptsize L$\sigma$M}}(s,t,u) & = & - 2\lambda \left( 1 -
\frac{2\lambda f^2_\pi}{m^2_\sigma - s} \right) \nonumber \\
& = & \frac{m^2_\sigma - m^2_\pi}{m^2_\sigma - s}\;\frac{s-m^2_\pi}{f^2_\pi}\;,
\end{eqnarray}
so that the $I=0\,$ S-wave amplitude $3 A + B + C$ generates a $23\%$
{\em enhancement} \/of Eq.~(\ref{eqweinsc1}):
\begin{eqnarray}
\left.a_{00}\right|_{\mbox{\scriptsize L$\sigma$M}} & = &
\frac{7+\varepsilon}{1-4\varepsilon}\frac{m_\pi}{32 \pi f^2_\pi} +
{\cal O}(\varepsilon^2) \nonumber \\
& \simeq & 1.23 \; \frac{7}{32\pi} \frac{m_\pi}{f^2_\pi} \; \simeq \; 0.20\;
m^{-1}_{\pi}\; ,
\label{eqlsmsc1}
\end{eqnarray}
for $s=4 m^2_\pi$, $t=u=0$, $\varepsilon = m^2_\pi/m^2_\sigma \simeq 0.045$.
Equation~(\ref{eqlsmsc1}) is compatible with the recently measured ($K_{e4}$)
E865 data \cite{Pislak:2001bf}
\begin{equation} a_{00} \; = \; 0.216\pm 0.013 \; m^{-1}_\pi \; .
\end{equation}

However two-loop ChPT in Refs.~\cite{Ananthanarayan:2004xy,Pelaez:2003eh}
predicts $a_{00} = 0.22\,m^{-1}_\pi$ or $a_{00} = 0.23\,m^{-1}_\pi$, using Roy
equations instead of the L$\sigma$M. But we suggest
Eqs.~(\ref{eqweinsc1}--\ref{eqlsmsc1}) tell the whole L$\sigma$M story (and
ChPT with very many parameters adds nothing new). We would like to stress that
--- even though Eq.~(\ref{eqlsmsc1}) has been derived without taking into
account $t$-channel vector-meson-exchange contributions --- the result will
change at most very slightly when including such contributions, as they are
``chirally shielded'' in $I=0\,$ S-wave $\pi\pi$ scattering close to the
$\pi\pi$ threshold, due to the presence of an extra contact term
\cite{Harada:1995dc}.

It is worth pointing out that Weinberg's result Eq.~(\ref{eqweinsc1}),
amounting to leading-order non-unitarized ChPT, is obtained from the
L$\sigma$M result Eq.~(\ref{eqlsmsc1}) by taking the limit
$m_\sigma\rightarrow \infty$. This limit essentially implies a
\textit{nonlinear}-\/$\sigma$-model framework, which unfortunately has been
dominating theoretical descriptions of $\pi\pi$ scattering during the past 
three decades, despite its technical complications, deviations from
experimental data, and the mounting experimental evidence for a light
scalar-meson nonet.
\section{\bom{\gamma\gamma\rightarrow\pi^0\pi^0}}
\label{sec10}
Assuming that the $S$-wave process $\gamma\gamma\to\pi^0\pi^0$ proceeds via an
intermediate $\sigma$ resonance ($\gamma\gamma\to\sigma\to\pi^0\pi^0$), with
$m_\sigma \simeq 700$~MeV, Ref.~\cite{Kaloshin:1986yy} anticipated that the
cross section would be very small (\mbox{$< 10$~nb}). In fact, this was later
confirmed \cite{Marsiske:1990hx} with Crystal-Ball data. Then, in 1991,
Ref.~\cite{Ivanov:1991da} studied
$a_1\to\pi(\pi\pi)_{\mbox{\scriptsize $S$-wave}}$, and used the Dirac identity
\begin{equation}
\frac{1}{\not\!p - m} \, 2m\gamma_5 \, \frac{1}{\not\!p - m} = -\,\gamma_5 \,
\frac{1}{\not\!p - m} - \frac{1}{\not\!p - m} \,\gamma_5
\end{equation}
to verify that the sum of the quark-box and -triangle diagrams vanishes. The
same holds for $\gamma\gamma\to\sigma\to\pi^0\pi^0$ \cite{Babukhadia:1999xw}.
The amplitude for the latter process would even vanish exactly in the CL.
In the \qllsm, this ``chiral shielding'', already mentioned above, has been
manifestly built in.

However, already in the abstract of the 1993 ChPT paper
Ref.~\cite{Dobado:1992zs}, it was stated that ``the one-loop [ChPT] computation
does {\em not} \/fit data, even close to threshold, because \textit{unitarity}
effects are important, even at very low energies''. In effect, the latter
authors were arguing to unitarize ChPT, which later turned out
\cite{Oller:1998hw,Caprini:2005zr} to reinstate the broad $\sigma$ pole, and presumably even
recovers the QLL$\sigma$M theory. But then the nonperturbative effects of the
$\sigma$ meson with a finite mass would have to be taken into account, too,
in contrast with non-unitarized ChPT briefly discussed at the end of
Sec.~\ref{sec9}.

Then, in 1999, Ref.~\cite{Boglione:1998rw} suggested data leads to a
$\sigma\to\gamma\gamma$ decay rate
\begin{equation}
\Gamma(\sigma\rightarrow\gamma\gamma) = (3.8\pm 1.5) \; \mbox{keV} \; .
\end{equation}
Since this rate is given by
$m^3_\sigma|M_{\sigma\rightarrow\gamma\gamma}|^2/64\pi$, for
$m_\sigma = 650$~MeV the amplitude magnitude equals \cite{Scadron:2003yg}
\begin{equation}
|M_{\sigma\rightarrow\gamma\gamma}| \; = \; (5.3\pm 1.0) \times 10^{-2} \;
\mbox{GeV}^{-1} \; .
\label{eqmsiggamgam1}
\end{equation}
In fact, the L$\sigma$M amplitude is (assuming here for simplicity a purely
nonstrange $\sigma$ meson) 
\begin{equation}
|M_{\sigma\rightarrow\gamma\gamma}|_{\mbox{\scriptsize L$\sigma$M}} \; \simeq
\; \frac{5}{3} \frac{\alpha}{\pi f_\pi} +\frac{1}{3} \frac{\alpha}{\pi f_\pi}
\; \simeq \; 5.0 \times 10^{-2} \; \mbox{GeV}^{-1} \; .
\label{eqlsmsiggg1}  
\end{equation}
Here, the dominant first term is the quark-loop analogue of the well-known
$\pi^0\to\gamma\gamma$ amplitude, while the small second term is due to meson
loops \cite{Karlsen:1993cm}. Note the excellent agreement between experiment
and \lsm\ theory (compare also the conclusions of Refs.~\cite{Schumacher:2005an}).

\section{Summary, Discussion, and Conclusions}
\label{sec11}
In the present paper, we have presented a side-by-side comparison of the 
\qllsm\ and standard ChPT, in the context of pion ChSB, on the respective
capability to reproduce a  
large variety of low-energy observables and other results. Concretely,
we have reviewed the pion and nucleon sigma terms, pion and kaon charge radii,
Goldberger-Treiman relations and discrepancies, the effective current quark,
pion and nucleon masses, the ground-state scalar $\bar{q}q$ nonet, the $I=0$
$\pi\pi$ scattering length, and the process $\gamma\gamma\to\pi^0\pi^0$. In
\em all \em cases, the \qllsm\ was shown to be clearly superior, in predictive
power as well as simplicity.
 
While these results should leave no doubt about the preferable scheme
to make predictions in low-energy strong-interaction physics, many people 
in the field will still object by claiming --- with inconclusive proof
\cite{Gasser:1983yg} and also ignoring very serious convergence problems
\cite{Hatsuda:1990tt} --- that ChPT \em is \em \/low-energy QCD, while alleging
that the \lsm\ is just a model or even ``unrealistic'' \cite{Gasser:1983yg}.

However, besides its demonstrated \cite{Elias:1984zh} linkage with infrared
QCD, the \qllsm\ is a good candidate for a theory of strong interactions \em
at high energies \em \/as well \cite{Kleefeld:2002au}, fulfilling the requirement of asymptotic
freedom, and displaying the same symmetries as QCD. This suprising
conclusion is the consequence of recent developments in mathematical physics,
proving that a meaningful quantum theory, with a bounded real spectrum, 
probability interpretation, and unitary evolution, is \textit{not}
restricted to \textit{Hermitian} \/Hamilton operators only. Instead,
the Hermiticity constraint can be relaxed to a weaker one,
called \textit{PT symmetry}
\cite{Bender:1998ke,PTworkshop:2003,PTworkshop:2004,PTworkshop:2005}, i.e.,
symmetry under parity and time-reversal transformations. This opens up the
possibility \cite{Kleefeld:2005hf} to construct an asymptotically free theory
of strong interactions on the basis of \textit{non-Hermitian} Hamilton
operators, not necessarily relying upon non-Abelian gauge fields, yet with
properties similar to QCD at high energies. At first sight, the idea of
allowing complex coupling constants in the \qllsm\ may seem far-fetched.
However, the upshot is that the bulk of the predictions of the QLLSM does not
depend on the (non-) Hermiticity of the Lagrangian, whereas certain observables
can \em only \em \/be understood if the quark-meson coupling is taken (close
to) imaginary. This might also provide a clue why quarks are not observed
on mass-shell at low energies, which is supported by very recent 
DSE \cite{Alkofer:2003jj} and lattice \cite{Bhagwat:2006tu} calculations. As
an example of a process requiring a non-Hermitian QLLSM Lagrangian, consider
the experimentally measured \em negative \em \/transition-form-factor ratio
$f_-^{K^+\!\pi^0}(0)/f_+^{K^+\!\pi^0}(0)=-0.125\pm0.023$
\cite{Eidelman:2004wy}, characterizing the semileptonic decay
$K^+\to\pi^0e^+\nu_e$, which can only be reproduced with an imaginary coupling
\cite{Kleefeld:2005hd}.

The obvious advantage of describing strong interactions with a generalized
\qllsm\ theory is its similarity \cite{Tornqvist:2005yp} with the mechanism
for spontaneous symmetry
breaking in the electroweak sector. This allows to treat strong and
electroweak interactions in many hadronic processes on an equal footing, and
with a minimum of parameters, in sharp contrast with ChPT. For
instance, the just mentioned decay $K^+\to\pi^0e^+\nu_e$ is then dominantly
described by a $W$-emission graph and a $\kappa$-exchange diagram
\cite{Kleefeld:2005hd}. From a more fundamental point of view, the manifest
finite divergence of the axial vector current underlying the QLL$\sigma$M
theory displays an instability of the effective action describing strong
interactions in the phase of broken chiral symmetry. It yields an instability
of strongly interacting Goldstone bosons of ChSB like the pion due to
electroweak decays. Nevertheless, the sum of the effective actions describing
strong and electroweak interactions must be stable. This imposes rigid
constraints on the overall sum of one-point functions, which will intimately
relate the parameters of strong and electroweak interactions to one another.

To conclude, it seems amazing that the Higgs scalar of the SMPP, with an
estimated mass of order $10^5$~MeV, takes responsibility for the tiny
nonstrange current quark masses and the ensuing nonvanishing pion mass, while
the scalar $\sigma$ meson with a mass of about $5\times10^2$~MeV
spontaneously generates considerably heavier dynamical quark masses, moreover
in such a way that the sum of the nonstrange current and dynamical quark
masses, i.e., the constituent quark masses, make up in the end practically
all of the proton's mass. Thus, let us make ours the words of E.\ Farhi and
R.\ Jackiw in the introduction of Ref.~\cite{Farhi:1982vt}:
\textit{``However, regardless of the source for weak-interaction
symmetry breaking, the Goldstone boson which is the longitudinal component
of the weak-interaction vector meson has a small $\approx f_\pi/300$~GeV
admixture of the QCD pion, and the physical pion has a small admixture of
the weak-interaction Goldstone boson.''}

\begin{acknowledgments}
We are indebted to E.~van Beveren for useful discussions and suggestions.
This work was supported by the {\it Funda\c{c}\~{a}o para a
Ci\^{e}ncia e a Tecnologia} \/of the {\it Minist\'{e}rio da
Ci\^{e}ncia, Tecnologia e Ensino Superior} \/of Portugal,
under contract POCTI/FP/FNU/50328/2003 and grant SFRH/BPD/9480/2002.
\end{acknowledgments}

\appendix

\section{The ``traditional'' \bom{U(3)\times U(3)} Linear \bom{\sigma} Model} 
\label{appendix1}
\subsection{Lagrangian, mass parameters, and mixing angles}
The Lagrangian of the ``traditional'' $U(3)\times U(3)$ L$\sigma$M before 
spontaneous symmetry breaking is given by \cite{Levy:1967a,Gasiorowicz:1969kn,
Cabibbo:1970uc,Delbourgo:1998kg,Kleefeld:2005qs}
\begin{eqnarray}
{\cal L}  & = & \frac{1}{2} \, \mbox{tr} [(\partial_\mu 
\Sigma_+)(\partial^\mu \Sigma_-)] - \frac{1}{2} \, \mu^2 \, \mbox{tr} [ 
\Sigma_+ \Sigma_-]\nonumber \\
  & - & \frac{\lambda}{2} \, \mbox{tr} [ \Sigma_+ \Sigma_- \Sigma_+ 
\Sigma_-]  - \frac{\lambda^\prime}{4} \, \Big( \mbox{tr} [ \Sigma_+ 
\Sigma_-]\Big)^2 \nonumber \\
  & + & \frac{\bar{\beta}}{2} \Big( \mbox{det} [ \Sigma_+] + \mbox{det} [ 
\Sigma_-]\Big) + \mbox{tr} [ C S] \; , \label{eqlagrlinsig1}
\end{eqnarray}
with $\Sigma_\pm (x) \equiv S(x) \pm i P(x)$ (see Eq.~(\ref{spmat})).
It is convenient to define also $\sigma_{n} \equiv (\sigma_{u\bar{u}} + 
\sigma_{d\bar{d}})/\sqrt{2}$, $\sigma_{3}\equiv (\sigma_{u\bar{u}} - 
\sigma_{d\bar{d}})/\sqrt{2}\simeq a^0_0$,
  $\eta_{n} \equiv (\eta_{u\bar{u}} + \eta_{d\bar{d}})/\sqrt{2}$, 
$\eta_{3}\equiv (\eta_{u\bar{u}} - \eta_{d\bar{d}})/\sqrt{2}\simeq \pi^0$.
Scalar and pseudoscalar meson masses are then determined after
(isospin-symmetric) spontaneous symmetry breaking $S\rightarrow S-D$, with 
$D\simeq\mbox{diag} (a,a,b)$, $f_\pi = \sqrt{2}a$, 
$f_K=(a+b)/\sqrt{2}$, $\beta\equiv - b\bar{\beta}$, and $X\equiv a/b$,
 by~\cite{Kleefeld:2005qs}
\begin{eqnarray}
m^2_{a_0} & = & \bar{\mu}^2 + 6\lambda a^2 + \beta\;, \nonumber \\
m^2_{\sigma_n} & = & \bar{\mu}^2 + (6\,\lambda + 4 \lambda^\prime ) a^2 - 
\beta \; , \nonumber \\
m^2_{\sigma_s} & = & \bar{\mu}^2 + (6\,\lambda + 2 \lambda^\prime ) b^2 \; 
, \nonumber \\
m^2_\kappa & = & \bar{\mu}^2 + 2\lambda (a^2+b^2+ab) + \beta X \; , 
\nonumber \\
m^2_\pi & = & \bar{\mu}^2 + 2\lambda a^2 - \beta \; , \nonumber \\
m^2_{\eta_n} & = & \bar{\mu}^2 + 2\lambda a^2 + \beta \; ,\nonumber \\
m^2_{\eta_s} & = & \bar{\mu}^2 + 2\lambda b^2 \; , \nonumber \\
m^2_K & = & \bar{\mu}^2 + 2\lambda (a^2+b^2-ab) - \beta X \; , 
\label{masslsm1}
\end{eqnarray}
with $\bar{\mu}^2 \equiv \mu^2 + \lambda^\prime (2a^2 + b^2)$, and
\begin{eqnarray}
  \sin(2\phi_P) & = & 2\sqrt{2}\,\beta X/(m^2_{\eta^\prime}-m^2_\eta) \, 
, \label{eqmixp1} \\
\sin(2\phi_S) & = & 2\sqrt{2}\,(-\beta X + 2\lambda^\prime 
a b)/(m^2_{f_0}-m^2_\sigma) \, , \label{eqmixs1}
\end{eqnarray}
where $\sigma = \sigma_n \cos \phi_S - \sigma_s \sin \phi_S$, $f_0 = 
\sigma_n \sin \phi_S + \sigma_s \cos \phi_S$, and  $\eta = \eta_n \cos 
\phi_P - \eta_s \sin \phi_P$, $\eta^\prime = \eta_n \sin \phi_P + \eta_s 
\cos \phi_P$. On the basis of Eqs.\ (\ref{masslsm1}) and the GTRs $f_\pi g 
= \hat{m}$ and $f_K g = (\hat{m}+m_s)/2$, it is straightforward to derive 
the L$\sigma$M predictions
\begin{eqnarray} m^2_{a_0} & = & m^2_{\eta_n} + 2 \, (\lambda/g^2) \, 
\hat{m}^2 , \label{eqmesmasspr1} \\
  m^2_{\sigma_n} & = &  m^2_\pi \; + 2 \, (\lambda/g^2) \; \hat{m}^2 \; 
(1+\lambda^\prime/\lambda) \; , \label{eqmesmasspr2}\\
m^2_{\sigma_s} & = & m^2_{\eta_s} + 2 \, (\lambda/g^2) \; m_s^2\; 
(1+\lambda^\prime/(2\,\lambda))\; ,\label{eqmesmasspr3} \\
m^2_{\kappa} & = & m^2_K + 2 \, (\lambda/g^2) \; \hat{m} \, m_s + 2\, 
\beta X\; .\label{eqmesmasspr4}
\end{eqnarray}
Inspired by the important relation $\lambda/g^2\simeq 2$,
which was obtained by Delbourgo and Scadron \cite{Delbourgo:1993dk} on
the basis of one-loop dynamical generation of the QLL$\sigma$M, we define
the coupling ratio $\xi\equiv \lambda/(2 g^2)\simeq 1$, which allows us,
also using Eq.~(\ref{eqmixp1}), to write
Eqs.~(\ref{eqmesmasspr1}--\ref{eqmesmasspr4}) in the more convenient form 
\begin{eqnarray} m^2_{a_0} & \simeq & \xi \; (2\hat{m})^2 + m^2_\eta \cos^2\phi_P 
+ m^2_{\eta^\prime} \sin^2\phi_P \, , \label{eqmesmassnew1} \\
  m^2_{\sigma_n} & \simeq & \xi \; (2\hat{m})^2 (1+\lambda^\prime/\lambda) + 
m^2_\pi \, ,\label{eqmesmassnew2} \\
m^2_{\sigma_s} & \simeq & \xi \; (2 m_s)^2 (1+\lambda^\prime/(2\lambda)) 
\nonumber \\
  & & + m^2_\eta \sin^2\phi_P + m^2_{\eta^\prime} \cos^2\phi_P 
,\label{eqmesmassnew3} \\
m^2_{\kappa} & \simeq & \xi \; (2 \hat{m})(2 m_s) \nonumber \\
  & & + m^2_K  + (m^2_{\eta^\prime}-m^2_\eta) \sin(2\phi_P)/\sqrt{2}\, . 
\label{eqmesmassnew4}
\end{eqnarray}
Note that Eqs.\ (\ref{masslsm1}) unambiguously imply for the mass 
parameter of the $\kappa$ meson in the $U(3)\times U(3)$ L$\sigma$M
\begin{equation} m^2_\kappa \; = \; \frac{(f_K/f_\pi)\, m^2_K - 
m^2_\pi}{(f_K/f_\pi) - 1} \;\simeq \; (1128\; \mbox{MeV})^2 \; ,
\label{eqkappamass1}
\end{equation}
with $f_K/f_\pi\simeq 1.22$, and the isospin-averaged masses
$m_{\pi}=138.0$~MeV, $m_{K}=495.0$~MeV. Note that, as hinted already
in Sec.~\ref{sec8}, unitarization will then split such a single ``bare''
$\kappa$ state into the pair of physical resonances $\kappa$(800) and
$K_0^*$(1430) \cite{vanBeveren:2001kf}. Nevertheless, employing the value for
$m_\kappa$ from Eq.~(\ref{eqkappamass1}),
we may use Eq.~(\ref{eqmesmassnew4}), $m_\eta=547.75$~MeV, 
$m_{\eta^\prime}=957.78$~MeV, and $m_s=1.44\times \hat{m}$ to determine 
the pseudoscalar mixing angle $\phi_P$ as a function of $\hat{m}$:
\begin{equation} \sin(2\phi_P) \simeq \sqrt{2} \; \frac{m^2_{\kappa} - 
m^2_K - \xi \; (2 \hat{m})(2 m_s)}{m^2_{\eta^\prime}-m^2_\eta}  \; 
.\label{eqphipmhat1}
\end{equation}
The next step is to determine the scalar mixing angle $\phi_S$ as a 
function of $\hat{m}$ and some given value of the coupling ratio 
$\lambda^\prime/\lambda$ (which is experimentally known to be very small
\cite{Kleefeld:2005qs}), via the identity
\begin{equation}\tan(2\phi_S) \simeq \frac{\sqrt{2}\;\xi \,(2\hat{m})(2m_s) 
\displaystyle\frac{\lambda^\prime}{\lambda} - \sin(2\phi_P) 
(m^2_{\eta^\prime}-m^2_\eta)}{m^2_{\sigma_s}-m^2_{\sigma_n}} \,,
\end{equation}
where $m^2_{\sigma_n}$ and $m^2_{\sigma_s}$ are given by
Eqs.~(\ref{eqmesmassnew2}) and (\ref{eqmesmassnew3}), respectively.
Given $\phi_S$, we can finally write $m_\sigma$ and $m_{f_0}$ as
\begin{eqnarray}
m^2_\sigma & = & \frac{1}{2}\left(m^2_{\sigma_s}+m^2_{\sigma_n} - 
\frac{m^2_{\sigma_s}-m^2_{\sigma_n}}{\cos(2\phi_S)}\right) \; ,
\label{eqsigmix1a} \\
m^2_{f_0}  & = & \frac{1}{2}\left(m^2_{\sigma_s}+m^2_{\sigma_n} + 
\frac{m^2_{\sigma_s}-m^2_{\sigma_n}}{\cos(2\phi_S)}\right) \; .
\label{eqsigmix2a} 
\end{eqnarray}
The importance of the aforementioned ``traditional'' $U(3)\times U(3)$ \lsm\
lies in the fact that it has many features of a yet to be determined
effective action, constructed on the basis of the \qllsm\ by integrating out
quarks and disregarding vector and axial-vector mesons. Furthermore, it
provides a mechanism that {\em simultaneously} --- and almost quantitatively
--- explains mixing for pseudoscalar {\em as well as} \/for scalar mesons.
Empirically we notice that the model, with its limitations (no (axial)
vector mesons, no two-loop effects), prefers a coupling ratio
$\xi\simeq 0.9064$, for $X^{-1}=m_s/\hat{m}=1.44$ and $\hat{m}=337.5$~MeV, in
order to match the experimentally favored value $\phi_P =(41.2\pm 1.1)^\circ$
\cite{Kleefeld:2005qs,Kroll:2005sd} (compare also to
Refs.~\cite{Escribano:2005qq} and \cite{Bramon:1997mf}, for newer and older
experimental/theoretical findings for $\phi_P$, respectively), and
$\phi_S\simeq (- 18\pm 2)^\circ$~\cite{Kleefeld:2001ds,Kekez:2000aw}. On the
other hand,
the one-loop-generated value $\xi=1$ implies, by the same token, rather small,
almost chiral-limiting quark-mass values. In contrast with the theoretical
uncertainties in $\phi_S$ stemming from the mentioned limitations of the
``traditional'' $U(3)\times U(3)$ \lsm, there is at least one way to justify,
in a quite model-independent fashion, the theoretical predictions of the
``traditional'' $U(3)\times U(3)$ \lsm\ for $\phi_P$, to be explained next.

We recall the pseudoscalar mass matrix of $\eta_n$ and $\eta_s$, the
diagonalization of which led to Eqs.\ (\ref{eqmixp1}) and (\ref{eqphipmhat1}):
\begin{equation}
\left( \begin{array}{cc} m^2_\pi + 2\beta & \sqrt{2}  \beta X \\[1mm]
\sqrt{2}  \beta X & 2 m^2_K - m^2_\pi + \beta X^2 - (\beta + 2\lambda b^2)
(1-X)^2
\end{array} \right) \, .
\label{massmat1}
\end{equation}
For $(\beta + 2\lambda b^2) (1-X)^2=0$, this reduces to the mass matrix
\begin{equation} \left( \begin{array}{cc} m^2_\pi + 2\beta & \sqrt{2}  \beta X
\\[1mm] \sqrt{2}  \beta X & 2 m^2_K - m^2_\pi + \beta X^2 \end{array}\right)\,,
\label{massmat2}
\end{equation}
motivated and discussed in Refs.~\cite{Jones:1979ez,Kekez:2000aw,Kekez:2002uh},
the diagonalization of which implies in a ``model-independent'' way: 
\begin{eqnarray}
\beta&=&\frac{(m_{\eta^\prime}^2-m^2_\pi)(m_\eta^2-m^2_\pi)}{4(m^2_K-m^2_\pi)}
\simeq 0.2792\:\mbox{GeV}^2 ,
\label{modindep1} \\[1mm]
X&=&\sqrt{\frac{2\,(m_{\eta^\prime}^2-2m^2_K+m^2_\pi)(2m^2_K-m^2_\pi-m_\eta^2)}
{(m_{\eta^\prime}^2-m^2_\pi)(m_\eta^2-m^2_\pi)}} \nonumber \\[1mm]
& \simeq & 0.7776 \;=\; 1/1.2860\, , \label{modindep2} \\[1mm]
\phi_P&=&\arctan\sqrt{\frac{(m_{\eta^\prime}^2-2m^2_K+m^2_\pi)
(m_\eta^2-m^2_\pi)}{(2m^2_K-m^2_\pi-m_\eta^2)(m_{\eta^\prime}^2-m^2_\pi)}} \nonumber \\[1mm]
 & \simeq & 42.1^\circ \, .
\label{modindep3}
\end{eqnarray}
Notice that this $\phi_P$ is fully compatible with the optimal experimental
value of $(41.2\pm1.1)^\circ$.
Recalling now the definitions
$\eta_0\equiv(\eta_{u\bar{u}}+\eta_{d\bar{d}}+ \eta_{s\bar{s}})/\sqrt{3}$ and
$\eta_8\equiv (\eta_{u\bar{u}}+\eta_{d\bar{d}}-2\eta_{s\bar{s}})/\sqrt{6}$,
we conclude, using Eqs.~(\ref{massmat2}), (\ref{modindep1}), (\ref{modindep2}),
and (\ref{modindep3}), with the following ``model-independent'' predictions:
\begin{eqnarray}
m^2_{\eta_n} & = & m^2_\pi+2\beta\;\simeq\;(760.0\;\mbox{MeV})^2\;,
\nonumber \\ \nonumber \\[-3mm]
m^2_{\eta_s} & = & 2m^2_K-m^2_\pi+\beta X^2\;\simeq\;(799.9\;\mbox{MeV})^2\; ,
\nonumber \\ \nonumber \\[-3mm]
m^2_{\eta_0} & = & \frac{2m_K^2 + m^2_\pi}{3} + \frac{\beta(X+2)^2}{3} \;
\simeq \; (942.2\; \mbox{MeV})^2\; , \nonumber \\
m^2_{\eta_8} & = & \frac{4m_K^2 - m^2_\pi}{3} + \frac{2\beta(X-1)^2}{3} \;
\simeq \; (574.1\; \mbox{MeV})^2\, . \nonumber \\ \label{modindep4}
\end{eqnarray}

Note that indeed \cite{Kekez:2000aw} $m^2_{\eta_n}+m^2_{\eta_s}=1.21744$ GeV$^2
\approx m^2_{\eta_0}+m^2_{\eta_8}=1.21733$ GeV$^2
\approx m^2_{\eta}+m^2_{\eta'}=1.21737$ GeV$^2$, as should be in our mixing
scheme employing squared masses.

\subsection{Remark on the CL of the ``traditional'' \bom{U(3)\times U(3)}
L\bom{\sigma}M}
In the CL, $m_\pi \rightarrow m^{\cl}_\pi=0$ and $m_K\rightarrow m^{\cl}_K=0$
should hold. Hence, the ``traditional'' $U(3)\times U(3)$ 
L$\sigma$M predicts, due to Eq.~(\ref{eqkappamass1}), in the CL
\begin{equation}
(m^{\cl}_\kappa)^2 \left(\frac{f^{\cl}_K}{f^{\cl}_\pi} 
-1\right) \; = \; \frac{f^{\cl}_K}{f^{\cl}_\pi} \; (m^{\cl}_K)^2 - 
(m^{\cl}_\pi)^2 \; \stackrel{\displaystyle!}{=} \; 0 \; ,
\end{equation}
implying $(\beta + 2\lambda b^2) (1-X)^2|_{\cl}=0$, i.e., either exact flavor
symmetry $f^{\cl}_\pi=f^{\cl}_K$
($\Leftrightarrow \hat{m}^{\cl}=m^{\cl}_s\Leftrightarrow X_{\cl}=1$) or
$m^{\cl}_\kappa = 0$, which is not very desirable according to our foregoing
discussion. 

We thus learn that the ``traditional'' $U(3)\times U(3)$ L$\sigma$M ---
despite its success in describing mixing angles --- cannot be (without 
changes e.g.\ in the 't Hooft determinant, or further extensions like the 
inclusion of (axial-)vector mesons) the effective $U(3)\times U(3)$ 
meson L$\sigma$M that would result from the QLL$\sigma$M by integrating out 
quarks.

Fortunately, we know that the mass matrices Eq.~(\ref{massmat1}) and
(\ref{massmat2}) describing the $\eta-\eta^\prime$ system are rather
insensitive to likely changes. Hence, their predictions remain reliable
even in the CL, where they reduce to
\cite{Jones:1979ez,Kekez:2000aw,Kekez:2002uh}
\begin{equation}
\left.\left( \begin{array}{cc}
2\beta & \sqrt{2}  \beta X \\[1mm] \sqrt{2} \beta X & \beta X^2
\end{array} \right)\right|_{\cl} =
\beta_{\cl} \left( \begin{array}{c}
\sqrt{2} \\[1mm] X_{\cl}
\end{array} \right) (\sqrt{2}\;, \; X_{\cl}) \, , \\[2mm]
\end{equation}
with eigenvalues $(m^{\cl}_\eta)^2,(m^{\cl}_{\eta^\prime})^2\in
\{ 0, \beta_{\cl} (2+X_{\cl}^2)\}$.

\end{document}